\def\kms{km$\cdot$s$^{-1}$}
\def\nlff{{\em nlff}}

\documentclass{aa}  
\usepackage{natbib}
\usepackage{graphicx}
\usepackage{txfonts}
\begin{document}
   \title{Coronal Alfv\'en speeds in an isothermal atmosphere \\
   I. Global properties}
   
   \author{S. R\'egnier
          \and
          E. R. Priest
	  \and
	  A. W. Hood
          }
   \authorrunning{R\'egnier et al.}	

   \offprints{S. R\'egnier}

   \institute{School of Mathematics, University of St Andrews, St Andrews,
   Fife, KY16 9SS, UK\\
              \email{stephane@mcs.st-andrews.ac.uk}
             }

   \date{Received ; accepted }

 
  \abstract
   {}
   {Estimating Alfv\'en speeds is of interest in modelling the solar corona,
   studying the coronal heating problem and understanding the initiation and
   propagation of coronal mass ejections (CMEs). }
   {We assume here that the corona is in a magnetohydrostatic equilibrium and
   that, because of the low plasma $\beta$, one may decouple the magnetic
   forces from pressure and gravity. The magnetic field is then described by a
   force-free field for which we perform a statistical study of the magnetic
   field strength with height for four different active regions. The plasma
   along each field line is assumed to be in a hydrostatic equilibrium. As a
   first approximation, the coronal plasma is assumed to be isothermal with a
   constant or varying gravity with height. We study a bipolar magnetic field with a
   ring distribution of currents, and apply this method to four active
   regions associated with different eruptive events. }
   {By studying the global properties of the magnetic field strength above
   active regions, we conclude that (i) most of the magnetic flux is localized
   within 50 Mm of the photosphere, (ii) most of the energy is stored in the
   corona below 150 Mm, (iii) most of the magnetic field strength decays with
   height for a nonlinear force-free field slower than for a potential field.
   The Alfv\'en speed values in an isothermal atmosphere can vary by two orders
   of magnitude (up to 100000 km/s). The global properties of the Alfv\'en speed
   are sensitive to the nature of the magnetic configuration. For an
   active region with highly twisted flux tubes, the Alfv\'en speed is
   significantly increased at the typical height of the twisted flux bundles;
   in flaring regions, the average Alfv\'en speeds are above 5000 \kms~and
   depart strongly from potential field values.  }
   {We discuss the implications of this model for the reconnection rate and
   inflow speed, the coronal plasma $\beta$ and the Alfv\'en transit time.}


   \maketitle

\section{Introduction}
\label{sec:intro}

In \citeyear{alf42}, \citeauthor{alf42} discovered that magnetohydrodynamic
(MHD) waves are of three kinds: two magnetoacoustic waves (slow and fast modes)
which are compressible and can be subject to damping, and the so-called
Alfv\'en wave which propagates along the magnetic field and is incompressible.
The Alfv\'en wave propagates at the Alfv\'en speed given
by:
\begin{equation}
v_A = \frac{B}{\sqrt{\mu_0 \rho}}
\end{equation}    
where $B$ is the magnetic field strength and $\rho$ is the density of the
plasma. Since Alfv\'en's work on MHD waves, the Alfv\'en speed has been shown
to be an important quantity in understanding many physical processes in the solar
atmosphere such as (i) the coronal heating problem, (ii) the (non-) equilibrium
state of the corona, (iii) the initiation and propagation of coronal mass
ejections (CMEs).

The coronal heating problem aims basically to identify and understand the
mechanisms responsible for sustaining the high temperature of the corona (about
1 MK) compared with the photosphere (about 4600 K). It is thought that the
heating mechanism for the corona is of magnetic origin: the magnetic energy
stored in the corona is released and converted into heat either by magnetic
reconnection or by dissipation of waves \cite[see review by][and reference
therein]{kli06}. The Alfv\'en speed appears as a tuning parameter in both
heating processes. For instance, in a reconnection model such as the Sweet-Parker
model \citep{swe58, par63}, the outflow speed is of the order the Alfv\'en speed
of the inflow region. Therefore, the measurement of the local Alfv\'en speed in
the corona is an estimate of the outflow speed in the case of a possible
reconnection process. Observationally, \citet{nar06} and \citet{nag06} have
analysed numerous EUV and soft X-ray flares to deduce an inflow speed of a few
tens of \kms~and a reconnection rate between 10$^{-3}$ and 10$^{-1}$ of the
Alfv\'en speed. These results are consistent with the estimates made by
\cite{der96}. In terms of MHD waves, the mode conversion between slow and fast
magnetoacoustic waves is the efficient when the sound speed is equal to the
Alfv\'en speed for a disturbance propagating along the magnetic field.
Nevertheless under coronal conditions, the heating by dissipation of Alfv\'en
wave is more likely than by magnetoacoustic waves. The dissipation of shear
Alfv\'en waves generated in the low corona can be considered as a plausible
heating mechanism especially when the dynamics of Alfv\'en waves is dominated by
phase-mixing \citep{hey83, hoo97, nak97} or resonant absorption \citep{gro88,
poe89, poe90}. The damping of shear Alfv\'en waves by phase-mixing is associated
with the existence of a gradient in the Alfv\'en speed and inhomogeneities in
the background stratification. Resonant absorption is an other viable mechanism
for coronal heating when the driving velocity of the magnetic structure is equal
to the local Alfv\'en speed. Recently \cite{mcl04, mcl06} showed that fast
magnetoacoustic waves can be dissipated near a null point in a low-$\beta$
plasma and that Alfv\'en waves are more likely dissipated near separatrices.
Both fast and Alfv\'en waves contribute to Joule heating in the corona localized
near topological elements. Furthermore, \cite{lon07a} showed how fast
magnetoacoustic waves maybe launched by transient reconnection. 

A coronal magnetic configuration can be considered to be in equilibrium if the
time scale of the evolution $\tau_{eq}$ is longer than both the Alfv\'en time
scale $\tau_{A}$ (or Alfv\'en transit time) and the reconnection time scale
$\tau_{rec}$ \citep[e.g.,][]{ant87, pri99}. For a typical length of 100 Mm and
an Alfv\'en speed of 2000 km$\cdot$s$^{-1}$, $\tau_{A}$ is of 1 min. The
reconnection time $\tau_{rec}$ is of few seconds. The transit time $\tau_{eq}$
is often assumed to be of the order of 10-20 min in the corona for a typical
length of 100 Mm and an upper limit of the plasma velocity of 100
km$\cdot$s$^{-1}$ corresponding to a sub-sonic coronal flow. In the limit of
$\tau_{eq} \gg \tau_{A} \gg \tau_{rec}$, the magnetic configuration can relax to
a minimum energy state, i.e. a linear force-free field according to
\cite{wol58}. Then the evolution of the magnetic field can be described as a
series of linear force-free equilibria \citep{hey84}. If $\tau_{eq} > \tau_{A}
\gg \tau_{rec}$, the magnetic configuration can relax to a nonlinear force-free
equilibrium. \cite{reg06} showed that the evolution of an active region can be
well described by a series of nonlinear force-free equilibria, the time span
between two equilibria being of 15 min.

In the different classes of CMEs, two main ingredients are often considered: the
existence of a twisted flux tube or sheared arcade that can store mass and
magnetic energy, and the existence of a current sheet above or below the twisted
flux tube or sheared arcade in order to change the connectivity of the field
lines by magnetic reconnection. Different models exist. The classical CSHKP
model \citep{car64, stu68, hir74, kop76} assumes the formation of a current
sheet below a rising twisted flux tube in the pre-flare phase leading to the 
eruption of the flux rope and the formation of post-flare loops. According to
\cite{car82} and contrary to \cite{kop76}, the evolution of post-flare loops is
better described by  slow-shock waves (ohmic heating). In the breakout model
\citep{ant99, mac04}, the increase of shear in the underlying field lines of a
quadrupolar configuration leads to field lines opening at the top of the
magnetic system. This model leads to the formation of a flux rope in 3D
\citep{lyn05}.  The development of sheared arcades and/or the instability of a
flux rope are then a prerequisite to a CME. The coronal sheared arcades are
formed by shearing motions of magnetic polarities in the photosphere. The
formation of unstable flux ropes and/or the destabilisation of an existing flux
rope in the corona are often considered: kink instability
\citep[e.g.,][]{hoo81}, flux emergence \citep[e.g.,][]{ama00, che00},
cancellation of flux \citep[e.g.,][]{par94, pri94}, tether-cutting
\citep[e.g.,][]{moo80}. As noticed above, the reconnection process is an
important step in the CME: the speed and the acceleration of a CME is related to
the local Alfv\'en speed and to the reconnection rate. Therefore, the nature of
CMEs can be inferred from the determination of the local Alfv\'en speed: to know
where the reconnection is more likely to happen and to estimate the speed of the
CME.

To our knowledge, few attempts have been made to characterise the Alfv\'en speed
in a coronal magnetic configuration. \cite{der96} performed a study of the
Alfv\'en speed in active regions, quiet-Sun regions and coronal holes based on
magnetic field estimates and/or potential extrapolations, the density and loop
length being derived from coronal observations. \cite{war05} derived the
Alfv\'en speed associated with the propagation of wave disturbances by modelling
the magnetic field of an active region by a magnetic dipole superimposed on that
of the quiet Sun as in \cite{man03} and constraining the density by
observations. Following \cite{lin02}, the local Alfv\'en speed in an isothermal
atmosphere with a constant gravity is assumed to decrease with height until
about 1.5 solar radii ($\sim$ 350 Mm above the surface) and then to increase.
When a solar wind component is added to the modelled atmosphere \citep{sit99},
\cite{lin02} showed that the Alfv\'en speed decreases with height but is still
consistent with Alfv\'en speed values in the low corona ($<$ 350 Mm). 
Nevertheless, the variation with height and the dynamical range of the local
Alfv\'en speed are not well-known in the low corona.     

As the physical processes mentioned above rely on the nature of magnetic
configurations in the corona, the first part of this paper is dedicated to a
determination of the magnetic field strength and its variation from the
photosphere to the corona. The coronal magnetic field is not easily measurable.
Several attempts have been made to measure the magnetic field in prominences
via the Hanle effect \citep{ler84}, or above active regions from gyro-emission
frequencies \citep{whi97}. Recently the possibility of measuring the magnetic
field based on infrared wavelengths has been investigated as well as the
inference of the coronal magnetic field strength from the properties of type II
bursts \citep{cho07}. \citet{pol75} derived the
magnetic field strength of several active region field lines using a
potential field extrapolation. Based on this method the magnetic field strength
decreases with height faster than $1/r$ from the photosphere and depends on the
total magnetic flux and the spatial distribution of polarities on the
photosphere. The significant results in measuring the coronal magnetic field
have been summarized by \citet{kou04}. Using a nonlinear force-free field
method to extrapolate the observed photospheric magnetic field into the corona,
we study the global properties of the magnetic field in the corona for
different active regions. In \cite{reg07a, reg07}, the authors described the
geometry of field lines as well as the variations of energy density with
height. In this paper, we focus on the distributions of the magnetic field
strength with height revealing different behaviours between different active
regions. 

The second part of this paper is to derive Alfv\'en speeds in the corona.
The assumption of our model is that the corona satisfies a magnetohydrostatic
equilibrium given by:
\begin{equation}
- \vec \nabla p + \rho \vec g + \vec j \wedge \vec B = \vec 0,
\label{eq:mhs}
\end{equation} 
describing the balance between plasma pressure $p$, gravity and magnetic
forces. Several magnetohydrostatic models already exist and are based on the
extrapolation of the magnetic field into the corona satisfying
Eqn.(\ref{eq:mhs}) with or without gravity \citep{low85, low86, low91, low92,
low93, neu95, low95, neu97}. Recently \cite{wie06a} have developed an
optimization scheme to address this problem. If we assume that the magnetic
forces dominate the corona, so that there is a force-free equilibrium for the
coronal magnetic field and a hydrostatic equilibrium for the distribution of the
plasma along each field line:
\begin{equation}
(\vec \nabla \wedge \vec B) \wedge \vec B = o(\beta)
\label{eq:ffo}
\end{equation}
and
\begin{equation}
\vec B \cdot \left(- \vec \nabla p + \rho \vec g\right) = 0.
\label{eq:hs} 
\end{equation}
These equations imply that there is no feedback of the coronal plasma on the
magnetic field. However the magnetic field strongly influences the plasma by
determining the shapes of coronal loops. Under coronal conditions, we assume
that the plasma $\beta$ is much less than 1: Eqn~(\ref{eq:ffo}) is a force-free
equilibrium and Eqn.~(\ref{eq:hs}) is the hydrostatic equilibrium. Solving these
equations, we investigate the values and distribution of the Alfv\'en speed in
the low corona assuming that the corona is  isothermal. We only consider the
global properties of the Alfv\'en speed and not the Alfv\'en speed profiles
along individual field lines. 

In Section~\ref{sec:obs_mod}, we first describe the observed active
regions, the magnetic field model as well as the isothermal atmosphere model
required to compute the Alfv\'en speeds. In Section~\ref{sec:model_param}, we then
study the field strength and Alfv\'en speed in a twisted bipolar field. We apply
this model to four solar active regions in Section~\ref{sec:ar}. We discuss the
implications of these results on the physics of active regions and CMEs in
Section~\ref{sec:disc}.

\section{Observations and models}
\label{sec:obs_mod}

	\subsection{Observed active regions} \label{sec:obs_ars}

To show the diverse and complex nature of the coronal magnetic field, we have
selected four different active regions with different types of activity
(confined flares, flares associated with a CME or filament eruptions) and at a
different stage of their evolution (before or after a flare):

\begin{itemize}
\item[-]{AR8151: observed on February 11, 1998 at 17:36 UT, this is an old
decaying active region (decreasing magnetic flux and magnetic polarities
diffusing away). A filament eruption associated with an aborted CME was
reported on Feb. 12, but no flare was observed. The vector magnetic field was
recorded by the MEES/IVM \citep{mic96, lab99}. The high values of the current
density imply strongly sheared and twisted flux bundles  \cite[see][]{reg02,
reg04}. Due to the existence of highly twisted flux tubes (with more than 1
turn) and the stability of the reconstructed filament and sigmoid (with less
than 1 turn), the authors concluded that the eruptive phenomena was most likely
to be due to the development of a kink instability in the highly twisted flux
bundles; }

\item[-]{AR8210: observed on May 1, 1998 from 17:00 to 21:30 UT, this is a
newly emerged active region with a complex topology as described in
\cite{reg06}. An M1.2 flare was recorded on May 1, 1998 at 22:30 UT. The
selected vector magnetogram (MEES/IVM) at 19:40 UT was observed during a 
``quiet'' period between two C-class flares. In \cite{reg06}, the authors 
described the magnetic reconnection processes occurring during this time period
and leading to a local reorganisation of the magnetic field. The reconnection
processes are related to the slow clockwise rotation of the main sunspot or a
fast moving, newly emerged polarity. Following the time evolution during 4
hours, the authors showed that the free magnetic energy decreases during 
the flare over a period of about 15 min, and the total magnetic energy is
slightly increased during this time period; }

\item[-]{AR9077: this corresponds to the famous Bastille day flare in 2000
\citep[e.g.,][]{liu01, yan01, fle01}. The vector magnetogram was recorded
at 16:33 UT after the X5.7 flare which occurred at 10:30 UT. The active region
was still in the magnetic reorganisation phase after the flare and ``post''-flare
loops were observed in 195\AA~TRACE EUV images. The flare was also associated
with a CME;}

\item[-]{AR10486: this active region is responsible for the main eruptions
observed during the Halloween events (26 Oct. to 4 Nov. 2003). The MEES/IVM
vector magnetogram was recorded on October 27, 2003 at 18:36 UT before the
X17.2 flare which occurred at 11:10 UT on October 28. The flaring activity of
this active region and the associated CMEs have been extensively studied. For
instance, \cite{met05} have shown that the large magnetic energy budget ($\sim
3 ~10^{33}$ erg) on Oct. 29 is enough to power the extreme activity of this
active region. }

\end{itemize}

For these particular active regions, the reduction of the full Stokes vector to
derive the magnetic field has already been detailed in several articles
\citep[e.g.,][]{reg02, reg06} -- the 180-degree ambiguity in the
transverse component was solved by using the algorithm developed in \cite{can93}
\citep[see also][]{met06}.

	\subsection{Nonlinear force-free field}

Under coronal conditions, the magnetic force dominates
the pressure gradient and gravity, and so we regard the corona above active
regions as being well described by the force-free approximation \citep[see e.
g., recent reviews by][]{reg07b, wie08}. Throughout this article, the coronal
magnetic configurations are computed from the nonlinear force-free (\nlff)
approximation based on a vector potential Grad-Rubin (\citeyear{gra58}) method
by using the XTRAPOL code \citep{ama97, ama99b}. The {\em nlff} field is
governed by the following equations:
\begin{equation}
\vec \nabla \wedge \vec B = \alpha \vec B,
\end{equation}
\begin{equation}
\vec B \cdot \vec \nabla \alpha = 0,
\label{eq:alphaline}
\end{equation}
\begin{equation}
\vec \nabla \cdot \vec B = 0,
\label{eq:divb}
\end{equation}
where $\vec B$ is the magnetic field vector in the domain $\Omega$ above the
photosphere, $\delta \Omega$, and $\alpha$ is a function of space defined as
the ratio of the vertical current density, $J_{z}$ and the vertical magnetic
field  component, $B_{z}$:
\begin{equation}
\alpha = \frac{1}{B_z}~\left( \frac{\partial B_y}{\partial x} - \frac{\partial
B_x}{\partial y} \right).
\end{equation} 
From Eqn.~(\ref{eq:alphaline}), $\alpha$ is constant along a field line. In
terms of the magnetic field $\vec B$, the Grad-Rubin iterative scheme can be
written in two steps as follows. Firstly,

\begin{equation}
\vec B^{(n)} \cdot \vec \nabla \alpha^{(n)} = 0 \quad \mathrm{in} \quad \Omega, 
\end{equation}
\begin{equation}
\alpha^{(n)}|_{\delta \Omega^{\pm}} = \alpha_{0},
\label{eq:bcalpha}
\end{equation}
where $\delta \Omega^{\pm}$ is defined as the domain on the photosphere for which
$B_{z}$ is positive ($+$) or negative ($-$). This step corresponds to the
transport of $\alpha$ along field lines, and secondly
\begin{equation}
\vec \nabla \wedge \vec B^{(n+1)} = \alpha^{(n)} \vec B^{(n)} \quad 
\mathrm{in} \quad \Omega,
\end{equation}
\begin{equation}
\vec \nabla \cdot \vec B^{(n+1)} = 0 \quad \mathrm{in} \quad \Omega, 
\end{equation}
\begin{equation}
B_{z}^{(n+1)}|_{\delta \Omega} = b_{z, 0},
\label{eq:bcbz}
\end{equation}
\begin{equation}
\lim_{|r| \rightarrow \infty}~|\vec B| = 0,
\label{eq:binf}
\end{equation}
which is the step required to update the magnetic configuration. The boundary
conditions on the photosphere are given by the distribution $b_{z, 0}$ of
$B_{z}$ on $\delta \Omega$ (see Eqn.~(\ref{eq:bcbz})) and by the distribution
$\alpha_{0}$ of $\alpha$ on $\delta \Omega$ for a given polarity (see
Eqn.~(\ref{eq:bcalpha})). We also impose that 
\begin{equation}
B_{n} = 0 \quad \mathrm{on} \quad \Sigma - \delta \Omega,
\label{eq:bcsurf}
\end{equation}
where $\Sigma$ is the surface of the computational box, $n$ refers to the
normal component to the surface. These conditions imply that no field line can
enter or leave the computational box, or in other words the active
region being studied is magnetically isolated. 

We ensure that the total unsigned magnetic flux is balanced within the
field-of-view by surrounding the vector magnetograms by SOHO/MDI line-of-sight
field. The smoothing function is applied in order to smooth out the transition
between the vector field and the MDI magnetogram. From the Grad-Rubin method, we
obtain a \nlff~equilibrium even if the photosphere is not exactly force-free
according to \citet{met95}. Recently, \citet{wie06} have developed a
pre-processing method in order to minimise the effects of non-force-freeness and
to obtain a more chromospheric magnetogram \citep{wie08a}. This way of improving
\nlff~field extrapolations is still under investigation. Our
\nlff~extrapolations have been successfully compared to EUV, soft X-rays and
visible observations \citep{reg02, reg04, reg06, reg04f, reg05c}. 

For the sake of comparison, we also compute the potential field with the same
boundary conditions. The potential field corresponds to the minimum energy state
that a magnetic configuration can reach if all the currents are dissipated and
while the distribution of the vertical field is held fixed.

	\subsection{Coronal Alfv\'en and sound speeds in an isothermal 
		atmosphere}
	\label{sec:model}

In order to compute the coronal Alfv\'en speeds, we first need to define the
thermodynamic parameters such as temperature, density and pressure.
As a first approximation, the corona is assumed to be an isothermal atmosphere
satisfying a hydrostatic equilibrium: 
\begin{equation}
-\vec \nabla p + \rho \vec g = \vec 0.
\end{equation}
In agreement with the Harvard-Smithonian model of the solar atmosphere, we
suppose the assumption is reasonable above a height ($z_0$) of about 5 Mm above
the photosphere. Let us then compare the effect of a constant gravity
($g_0$) with a gravitational field ($g(z)$) decreasing with distance from the
Sun. In an isothermal atmosphere, the pressure and density are either:
\begin{equation}
\centering
\left\{ \begin{array}{c l}
p(z) = p_0 \exp{\left(- \frac{z - z_0}{H}\right)}, & \quad \textrm{and} \\[0.2cm]
\rho(z) = \rho_0 \exp{\left(- \frac{z - z_0}{H}\right)} & \quad \textrm{for $g =
g_0$,}\\
\end{array}
\right.
\end{equation}
or
\begin{equation}
\centering
\left\{ \begin{array}{c l}
p(z) = p_0 \exp{\left(- \frac{R_{\odot}^2}{H~(R_{\odot} + z_0)} \left( \frac{z -
z_0}{R_{\odot} + z} \right) \right)}, & \quad \textrm{and} \\[0.2cm]
\rho(z) = \rho_0 \exp{\left(- \frac{R_{\odot}^2}{H~(R_{\odot} + z_0)} \left(
\frac{z - z_0}{R_{\odot} + z} \right)\right)} & \quad \textrm{for $g = g(z) =
\frac{g_0~R_{\odot}^2}{(R_{\odot} + z)^2}$,}\\
\end{array}
\right.
\end{equation}
where $H = k_BT/(\tilde{\mu}m_pg_0)$ is the pressure scale-height ($k_B =
1.38~10^{-23}$ J$\cdot$K$^{-1}$, $\tilde{\mu} = 0.6$ for a fully ionized coronal
plasma, $m_p = 1.67~10^{-27}$ kg and $g_0 = g(R_{\odot}) = 274$
m$\cdot$s$^{-2}$), $p_0$ and $\rho_0$ are characteristic values of the pressure
and density at $z_0$. The pressure scale-height and the density $\rho_0$ are the
two free parameters of the model. Typical values are: $H = 50$ Mm for a
temperature of 1 MK and $\rho_0 = 10^{9}$ cm$^{-3}$. 

In these models, the Alfv\'en and sound speeds are given by:
\begin{equation}
v_A(x, y, z) = \frac{B(x, y, z)}{\sqrt{\mu_0 \rho_0}} \exp{\left(\frac{z -
z_0}{2H}\right)},
\label{eq:va}
\end{equation}
for a constant gravitational field and
\begin{equation}
v_A(x, y, z) = \frac{B(x, y, z)}{\sqrt{\mu_0 \rho_0}} \exp{\left(
\frac{R_{\odot}^2}{2H~(R_{\odot} + z_0)} \left( \frac{z - z_0}{R_{\odot} + z}
\right)\right)},
\label{eq:vagz}
\end{equation}
for a gravitational field varying with height,
and
\begin{equation}
c_s^2 = \frac{\gamma p_0}{\rho_0} = \frac{k_BT}{\tilde\mu m_p},
\end{equation} 
where $B$ is the magnetic field strength and $\gamma=1$ in an isothermal
atmosphere. Therefore, the sound speed is constant in the corona and the
Alfv\'en speed can be defined locally knowing the magnetic field strength at
each point of the coronal volume. Assuming that at infinity the magnetic
field strength decays as $1/z^2$ for a dipolar field, we notice that
\begin{equation}
\lim_{z \rightarrow +\infty}~v_A(x, y, z) = + \infty,
\end{equation}  
for a constant gravitational field, while for a gravitational field decreasing
with height, we obtain
\begin{equation}
\lim_{z \rightarrow +\infty}~v_A(x, y, z) = 0.
\end{equation}
The assumption of a gravitational field varying with height is therefore more
reasonable at high altitudes in order to avoid the Alfv\'en speed becoming
unbounded and to have a smooth transition with the solar wind.

\section{Study of a bipolar field}
\label{sec:model_param}

	\subsection{Nonlinear force-free bipolar field}

Before applying this model to examples of solar active regions, we perform a
parameter study using a nonlinear force-free bipolar field. The free parameters
are the pressure scale-height (or temperature) and the density at the base of
the corona. 

\begin{figure}[!ht]
\centering
\includegraphics[width=.49\linewidth]{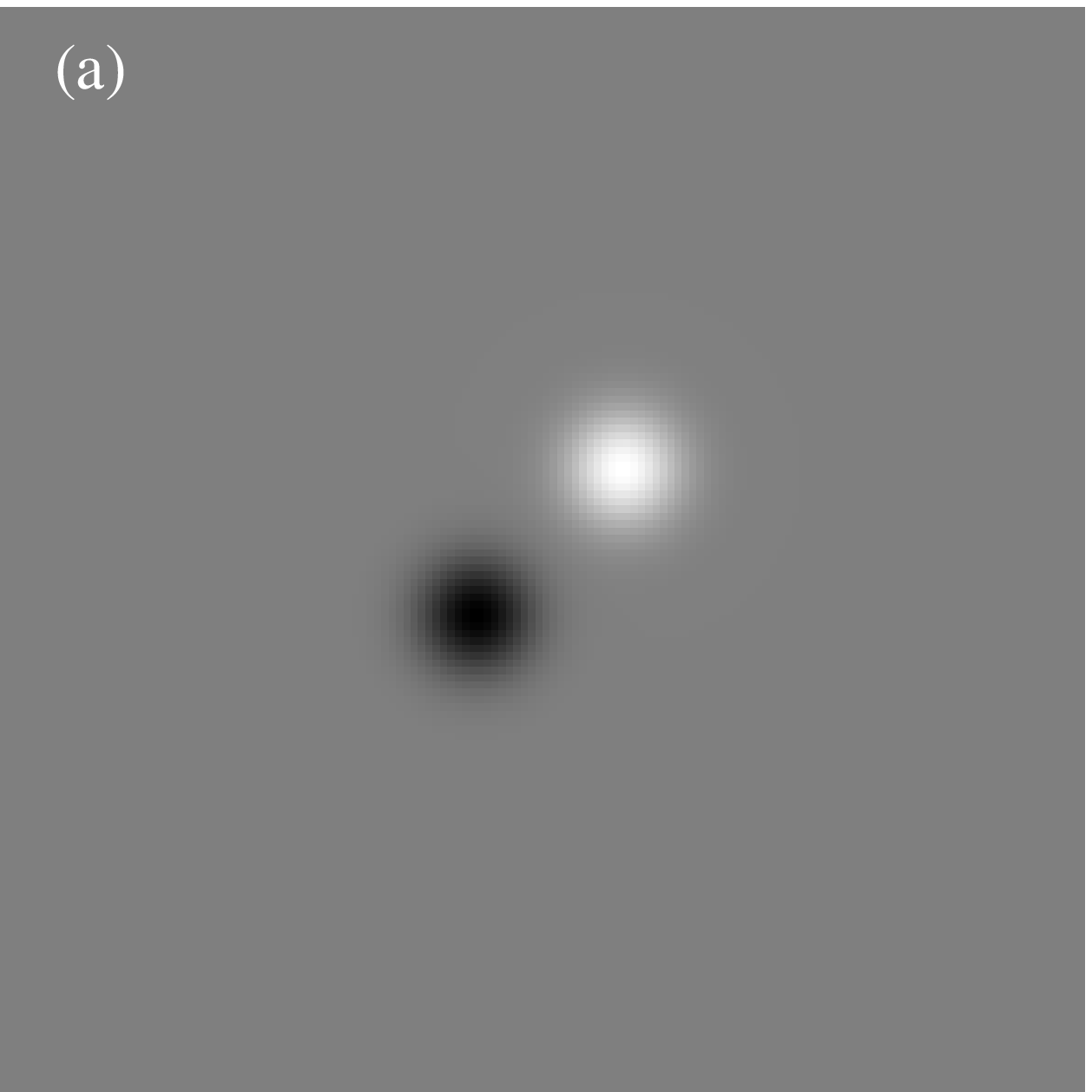}
\includegraphics[width=.49\linewidth]{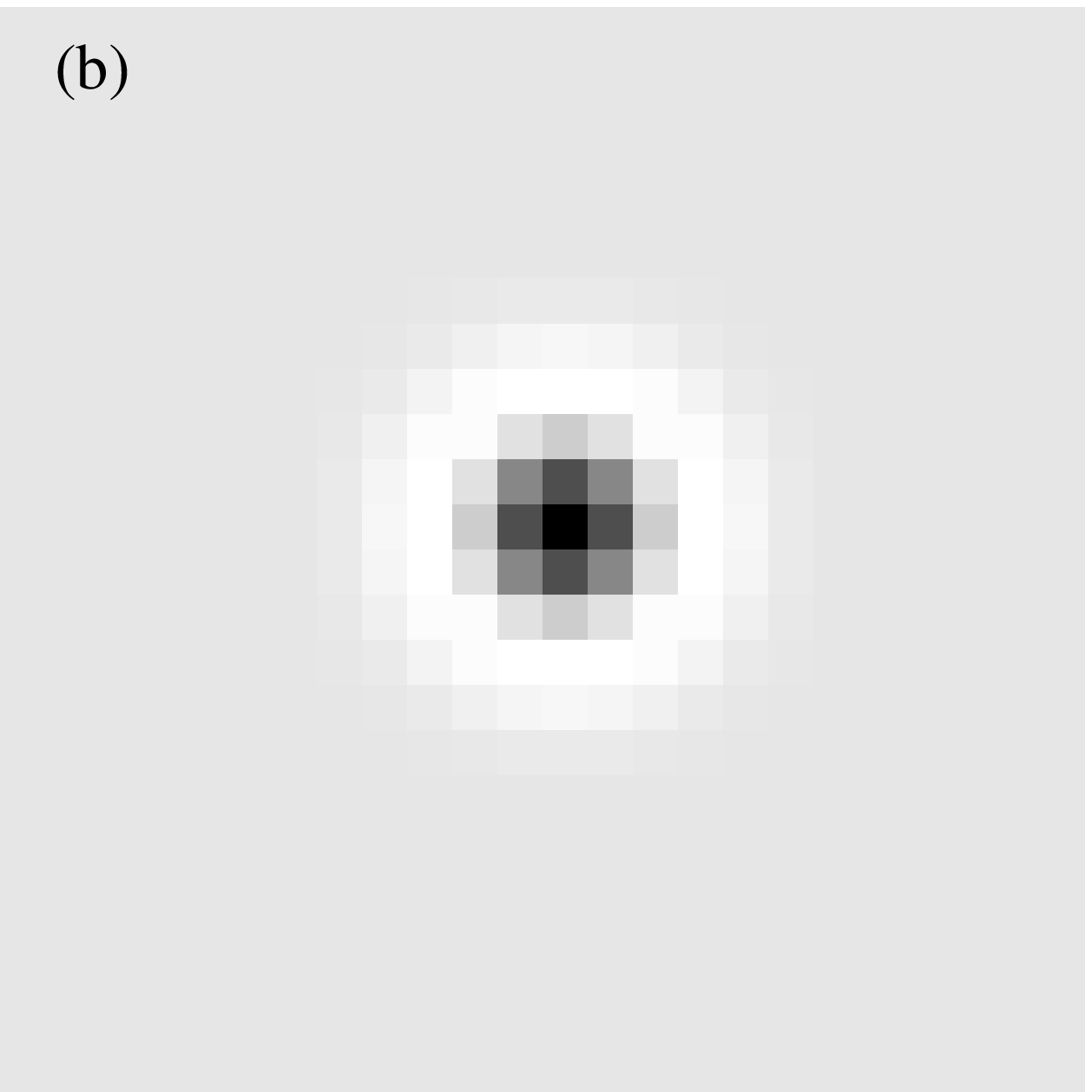}
\caption{(a) Vertical magnetic field on the photosphere, (b) vertical current
density distribution imposed in the positive polarity.}
\label{fig:dipmag}
\end{figure}

The bipolar vertical magnetic field component is simply given by a Gaussian
distribution with the same full-width at half-maximum for both polarities of 15
Mm (see Fig.~\ref{fig:dipmag}a). The maximum field strength is set to 2000 G.
The considered field-of-view has dimensions 150 Mm$\times$150 Mm assuming a
square pixel of 1 Mm$^{2}$. The separation between the two polarities is 50 Mm.
The magnetic flux is balanced and the total unsigned magnetic flux is 7.1
10$^{21}$ Mx. A typical bipolar magnetic field used in this study is depicted in
Fig.~\ref{fig:dipmag}a. The field-of-view is large enough to have no magnetic
flux on the edges of the surface. We impose a ring distribution of current at
the positive source of the form:
\begin{equation}
J_z(r) = 2~J_{z0}~[r^2 - C_0]~\exp{\left(-\frac{r^2}{\sigma^2}\right)},
\label{eq:rng}
\end{equation}
where $r$ is measured from the centre of the source and $C_0$ is a constant
which ensure a zero net current. We always take care to (i) keep a zero net
current, and (ii) have enough pixels to describe the steep gradient between the
positive and negative currents. A typical distribution used in this experiment
is shown in Fig.~\ref{fig:dipmag}b. We set $J_{z0}$ to 10 mA$\cdot$m$^{-2}$. To
satisfy the Grad-Rubin boundary-value problem, the current distribution is
defined in only one polarity. This equilibrium corresponds to a single twisted
flux tube with an excess magnetic energy of 3\% above the potential field.

	\subsection{Magnetic field strength} \label{sec:bdip}

\begin{figure}[!h]
\centering
\includegraphics[width=1.\linewidth]{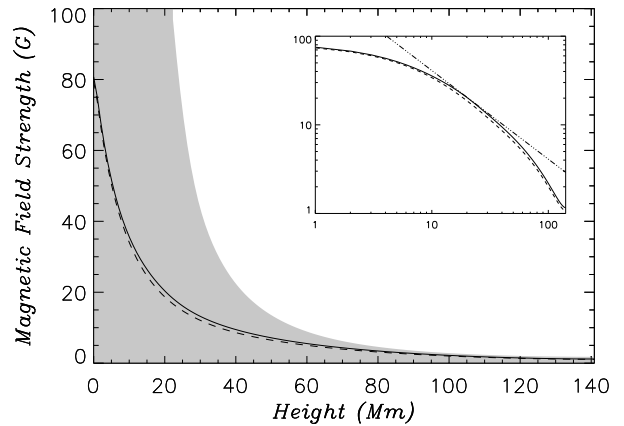}
\caption{Average field strength as a function of height (Mm) for the \nlff~field
(solid line) and the potential field (dashed line). The grey area is enclosed
between the minimum and maximum field strengths at a given height. The log-log
plot is drawn in the top-right corner. The dot-dashed line indicates the
$z^{-1}$ decay of the field.}
\label{fig:dip_bs}
\end{figure}

We first study the distribution of magnetic field strength with height for
potential and \nlff~fields. We plot the average field strength
as a function of height in Fig.~\ref{fig:dip_bs}. The average strength values
start from 80 G at the photosphere and tend to zero. The rapid decrease of the
field strength is due to the confinement of the twisted flux tube at a height
lower than 30 Mm due to a separation between polarities of 50 Mm. There is
a small departure of the \nlff~field values from the potential
values due to the storage of magnetic energy in the corona associated with the
twisted flux tube \citep[see e.g,][]{reg07}.
  
In Fig.~\ref{fig:dip_bs}, we also plot the log-log distribution of the average
field strength with height in order to determine the shape of the decaying
magnetic field. For both the potential and \nlff~fields, the average magnetic
field strength decays rapidly from the photosphere up to about 20 Mm and then
the rate of decline of the field strength reduces with a slope index greater
than -1 as indicated by the dot-dashed line in the log-log plot of
Fig.~\ref{fig:dip_bs}. We focus our interest on an index of -1 which corresponds
to the typical slope of a dipolar field average on a surface. The whole curve
is well fitted by a function of the following form:
\begin{equation}
\overline{B}(z) = B_0 \exp\left[-\ln^{2}\left(\frac{z}{z_0}\right)\right]
	= B_0 \left(\frac{z_0}{z}\right)^{\ln{(\frac{z}{z_0})}},
\end{equation} 
as deduced from the log-log plot, where $\overline B$ is the average field
strength and $z$ the height ($B_0$ and $z_0$ being constants). \citet{der96}
found that the magnetic field decay below 40 Mm is well described by an
exponential decay. We will discuss these typical curves further in
Section~\ref{sec:disc}.

	\subsection{Alfv\'en speed} \label{sec:model_va}

\begin{figure}[!h]
\centering
\includegraphics[width=1.\linewidth]{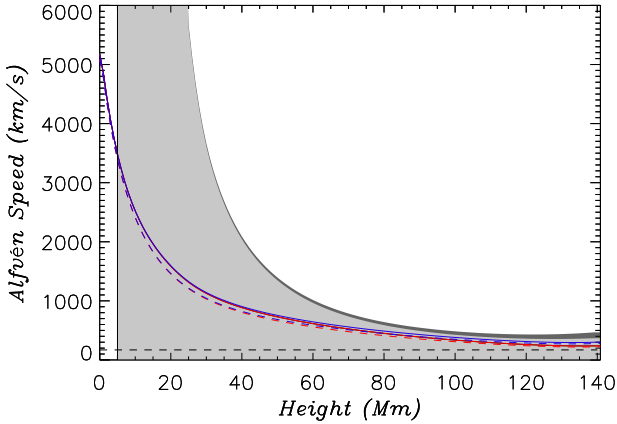}
\caption{Average Alfv\'en speed (\kms) as a function of height (Mm) for $H =$ 50
Mm and $\rho_0 = 10^{9}$ cm$^{-3}$. Blue and red solid curves are from the
\nlff~field for the constant and varying gravity models respectively. Blue and
red dashed curves are from the potential field for the constant and varying
gravity models respectively. The red and blue curves are very similar and almost
superimposed. The dark and light grey areas indicate the spread of Alfv\'en
speed values at a given height for the constant and varying gravity models
respectively: only above 60 Mm, one can slightly notice the difference between
the two gravity models.}
\label{fig:dip}
\end{figure}

\begin{figure}[!h]
\centering
\includegraphics[width=.49\linewidth]{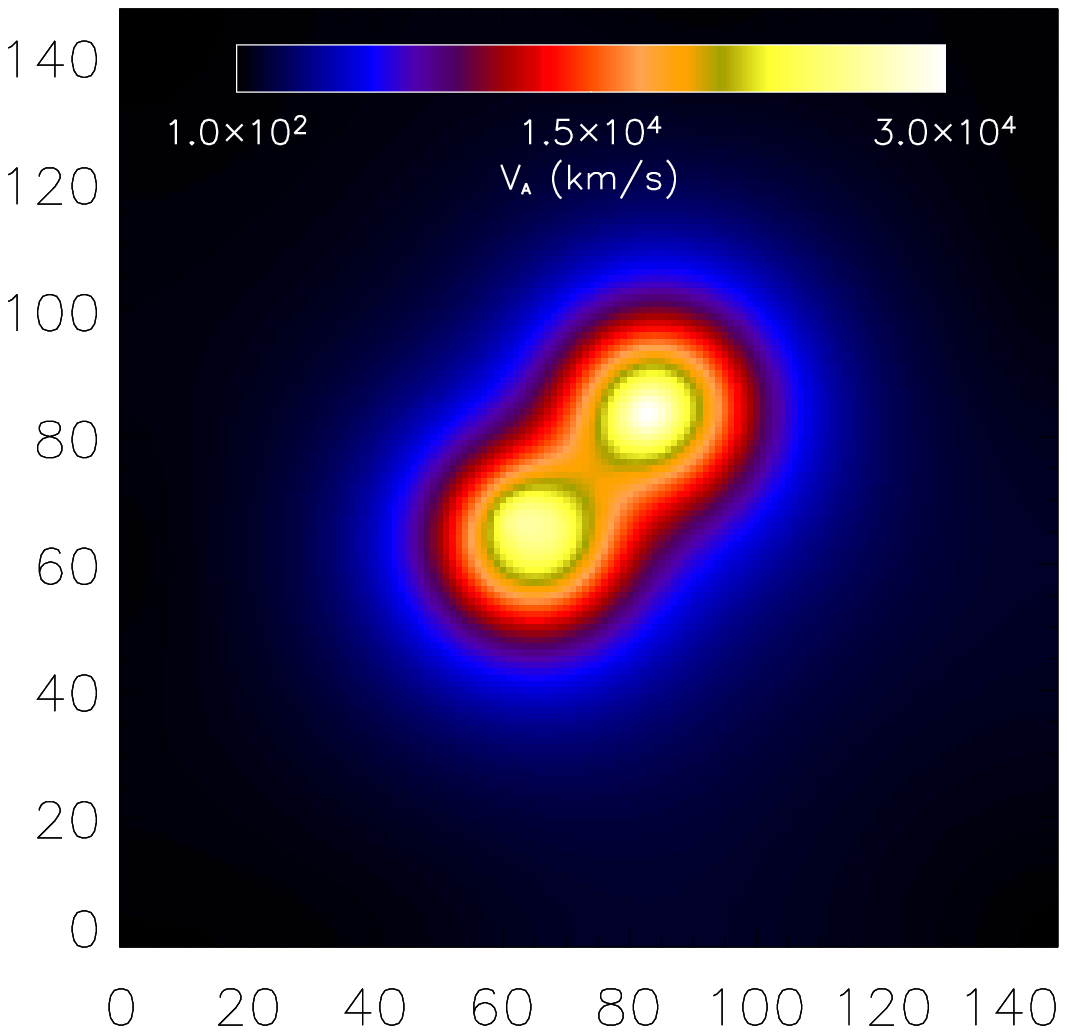}
\includegraphics[width=.49\linewidth]{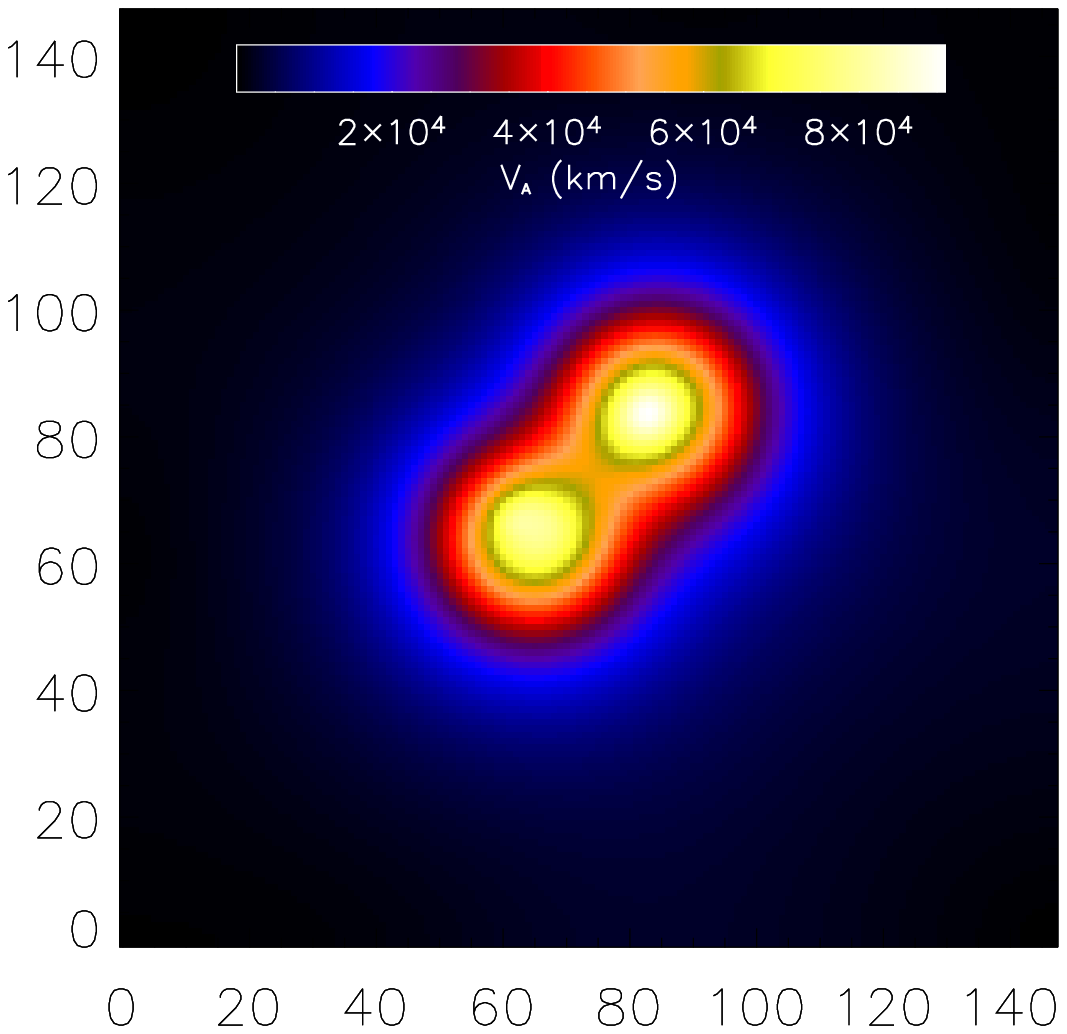}
\includegraphics[width=.49\linewidth]{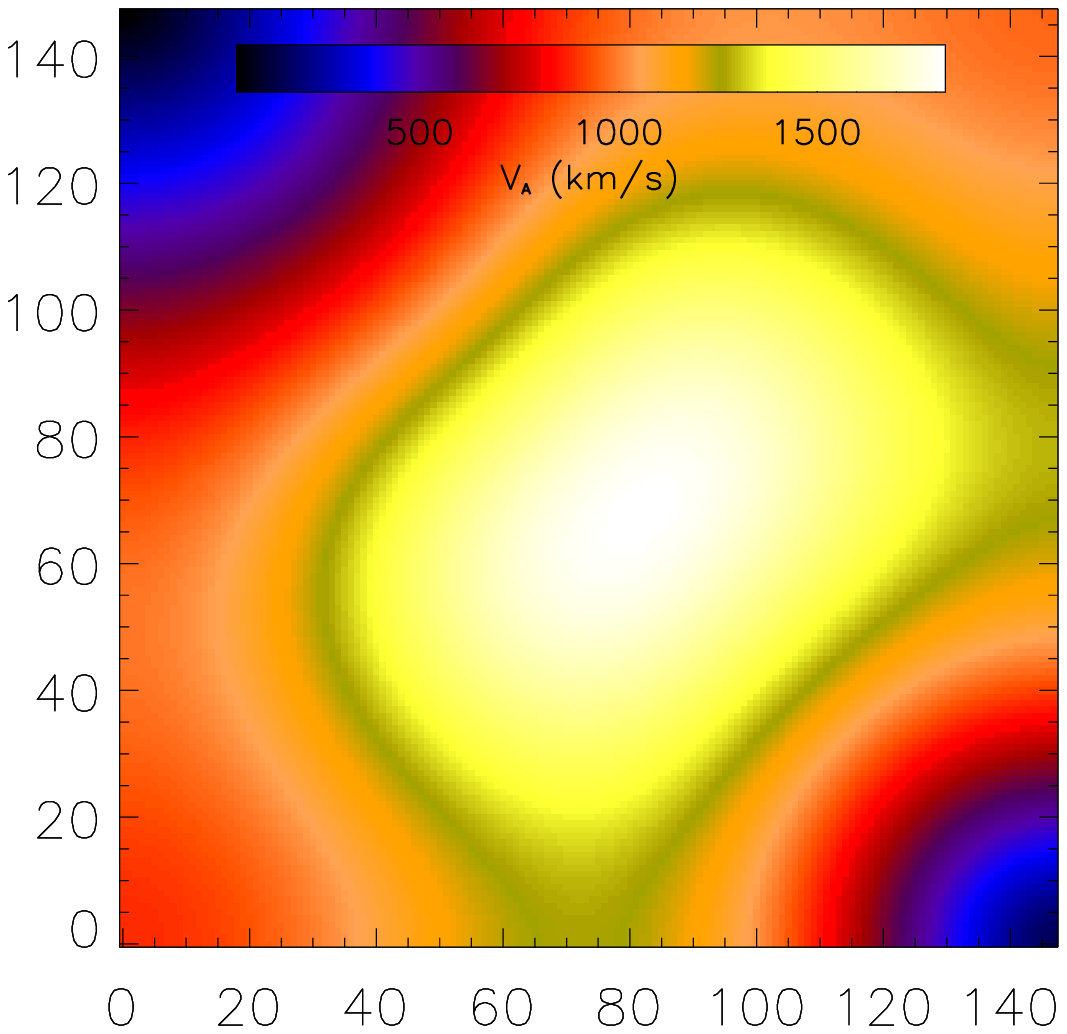}
\includegraphics[width=.49\linewidth]{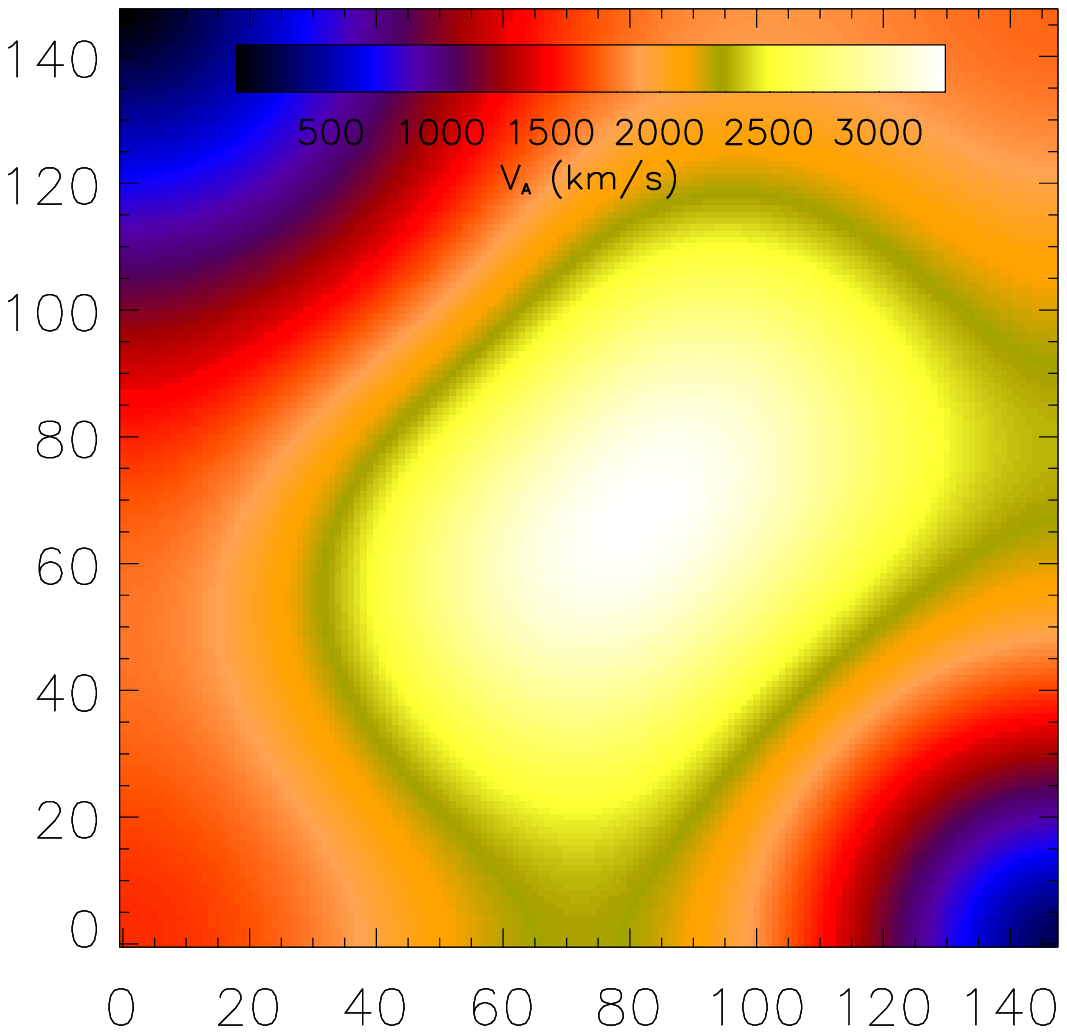}
\caption{Dynamic range of the Alfv\'en speed at  10 Mm (top row) and 60 Mm
(bottom row) for two sets of free parameters ($\rho_0$, $H$) in (cm$^{-3}$, Mm):
(10$^9$, 25), (10$^8$, 50) from left to right. }
\label{fig:bst}
\end{figure}

\begin{figure}
\centering
\includegraphics[width=1.\linewidth]{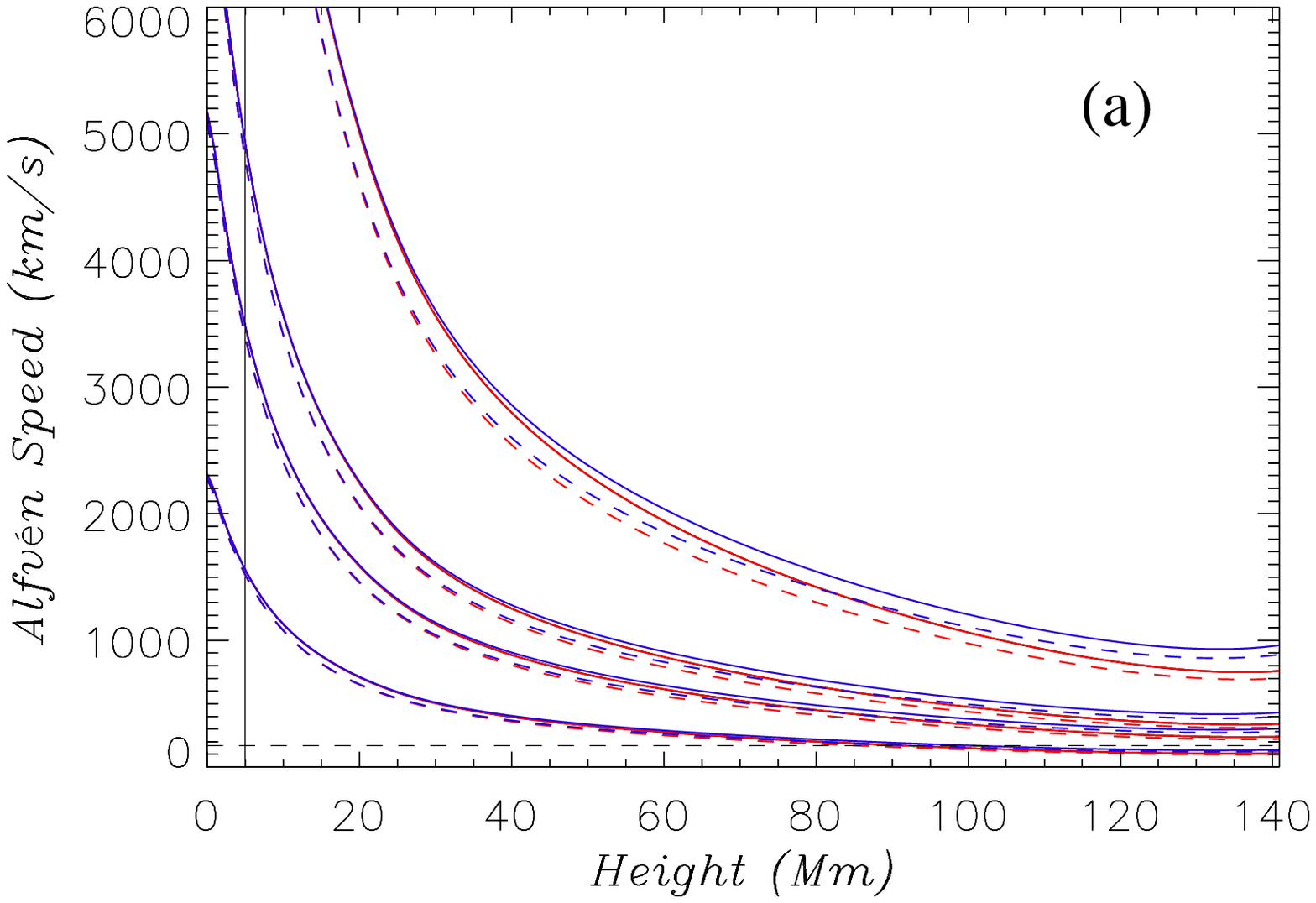}
\includegraphics[width=1.\linewidth]{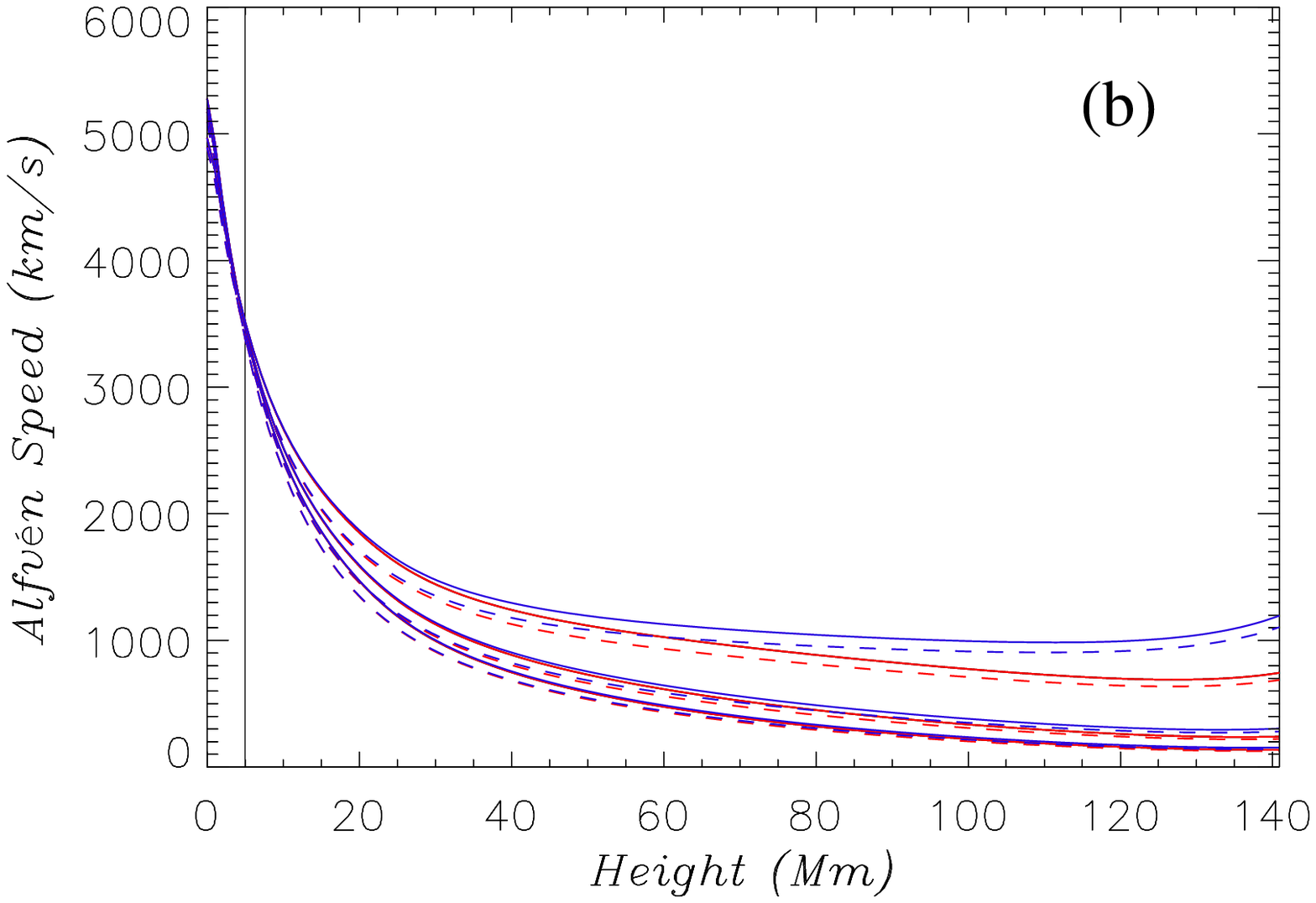}
\caption{Average Alfv\'en speed (km$\cdot$s$^{-1}$) as a function of height (Mm)
for the nonlinear force-free field (solid line) and the potential field (dashed
line). (a): for $H =$ 50 Mm and $\rho_0 = [5~10^{9}, 10^{9}, 5~10^{8}, 10^{8}]$
cm$^{-3}$ (from left to right); (b): for $H = [100, 50, 25]$ Mm (from left to
right) and $\rho_0 = 10^{9}$ cm$^{-3}$. Red (resp. blue) curves are for the
varying (resp. constant) gravity model.}
\label{fig:dip_rho0}
\end{figure}

We derive the Alfv\'en speeds from the model described above with $H =$ 50 Mm
(1 MK) and $\rho_0 =$ 10$^9$ cm$^{-3}$. In Fig.~\ref{fig:dip}, we plot the
average Alfv\'en speed at a given height from both the potential field (dashed
line) and the nonlinear force-free field (solid line). The Alfv\'en speed
values are valid only above 5 Mm (vertical solid straight line). Under the
force-free assumption, the plasma $\beta$ is considered to be much less than 1.
The plasma $\beta$ can be expressed in terms of the Alfv\'en and sound speeds:
$2 c_s^2/ \gamma v_A^2$. In an isothermal atmosphere, we then obtain a minimum
value for the Alfv\'en speed: 
\begin{equation}
v_{A, min}^2 = 2 c_s^2.
\end{equation} 
As the sound speed is 120 \kms~for $H = 50$ Mm, we plot $v_{A, min} =$ 170
\kms~in Fig.~\ref{fig:dip} and Fig.~\ref{fig:dip_rho0}a (dashed straight line). 
 The average Alfv\'en speed is similar for both gravity models in the case
of the bipolar field. The curves are exactly the same up to a height $z = H$.
Thus, we only report here the values for the constant gravity model. The average
Alfv\'en speed decreases with height suggesting that the magnetic field
decreases faster than the square root of the density with height. The Alfv\'en
speed values range between 2400 \kms~at 10 Mm and 300 \kms~at 140 Mm. The
scatter of the Alfv\'en speed (grey areas in Fig.~\ref{fig:dip}) strongly
depends on the height: a maximum value of 24000 \kms~at 10 Mm with a standard
deviation of 3800 \kms, and a maximum value of 990 \kms~at 60 Mm with a standard
deviation of 200 \kms~(see Fig.~\ref{fig:bst}).  By drawing the dynamic range of
the Alfv\'en speed at different heights (see Fig.~\ref{fig:bst}), we show that
the Alfv\'en speed is larger where the field strength is larger whatever the
density or the pressure scale-height: the two opposite polarities are clearly
seen at 10 Mm and the loop top is seen at 60 Mm.    

We now study the effects of two free parameters: the density at the base of the
corona $\rho_0$ (see Fig.~\ref{fig:dip_rho0}a) and the pressure scale-height
$H$ (see Fig.~\ref{fig:dip_rho0}b). In Fig.~\ref{fig:dip_rho0}a, we
plot the evolution of average Alfv\'en speed with height for several values of
the density $\rho_0$ varying from 10$^8$ to 5~10$^9$ cm$^{-3}$. As the magnetic
field is the same, the average Alfv\'en speed follows the density variations
which can be seen from the values of the Alfv\'en speed at 5 Mm: 1560
\kms~for $\rho_0 = 5~10^9$ cm$^{-3}$, 3500 \kms~for $\rho_0 = 10^9$ cm$^{-3}$,
4950 \kms~for $\rho_0 = 5~10^8$ cm$^{-3}$ and 11070 \kms~for $\rho_0 = 10^8$
cm$^{-3}$. We note that there is a minimum of these curves at about 135 Mm.
This minimum was also noticed by \cite{lin02} but at about 0.5 R$_{\odot}$
($\sim$ 350 Mm) for an isothermal atmosphere. From Eqn.~(\ref{eq:va}), the
minimum corresponds to the location where the variations of the gas pressure
are becoming more important than those of the magnetic field (that does not
mean that the plasma $\beta$ becomes greater than 1!). We conclude that this
minimum is strongly related to the distribution of the magnetic field: from
Eqn.~\ref{eq:va}, we deduce that the minimum of the curves is obtained for a
constant gravitational field when the magnetic field satisfies
\begin{equation}
\frac{\partial B}{\partial z} = - \frac{B}{2H},
\label{eq:min_va}
\end{equation}   
which depends only on the pressure scale-height. For a gravitational field
varying with height, the minimum occurs where
\begin{equation}
\frac{\partial B}{\partial z} = - \frac{B}{2H} \left( \frac{R_{\odot}}{R_{\odot}
+ z} \right)^2.
\label{eq:min_vagz}
\end{equation}   

We can already mention that
Eqn.~(\ref{eq:min_va}) is similar to the equation derived by \cite{fil00} for the
height of unstable prominences. In Fig.~\ref{fig:dip_rho0}b,
we plot the average Alfv\'en speed with height for several values of the
pressure scale-height varying from 25 to 100 Mm (from 0.5 to 2 MK, resp.)
giving a sound speed between 83 \kms~and 165 \kms, respectively. Therefore, the
Alfv\'en speed curve for $H =$ 100 Mm is less than $v_{A, min}$ above 100 Mm,
and the curve for $H =$ 25 Mm is way above $v_{A, min}$ whatever the height.
The minimum of the curves is decreasing with height when $H$ decreases. The
Alfv\'en speed values increase significantly when $H$ decreases: at 100 Mm, the
Alfv\'en speed is 200 \kms~for $H =$ 100 Mm, 400 \kms~for $H =$ 50 Mm and 1000
\kms~for $H =$ 25 Mm.

From this study, we conclude that a density at the base of the corona of 10$^9$
cm$^{-3}$ and a pressure scale-height of 50 Mm are a reasonable choice of
parameters to model an isothermal corona. We will adopt these values in the
following Sections. 

\begin{figure*}[!ht]
\centering
\includegraphics[width=.49\textwidth]{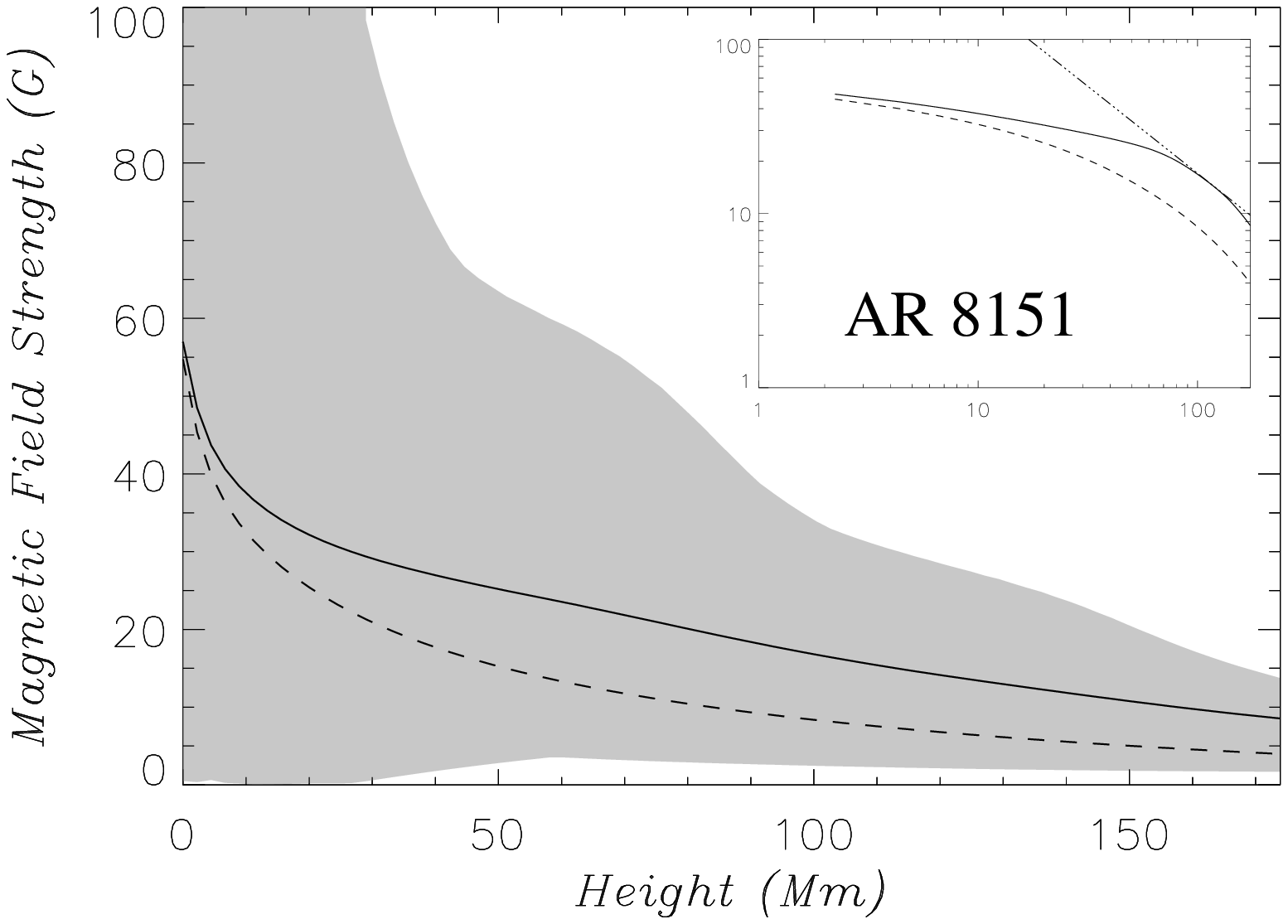}
\includegraphics[width=.49\textwidth]{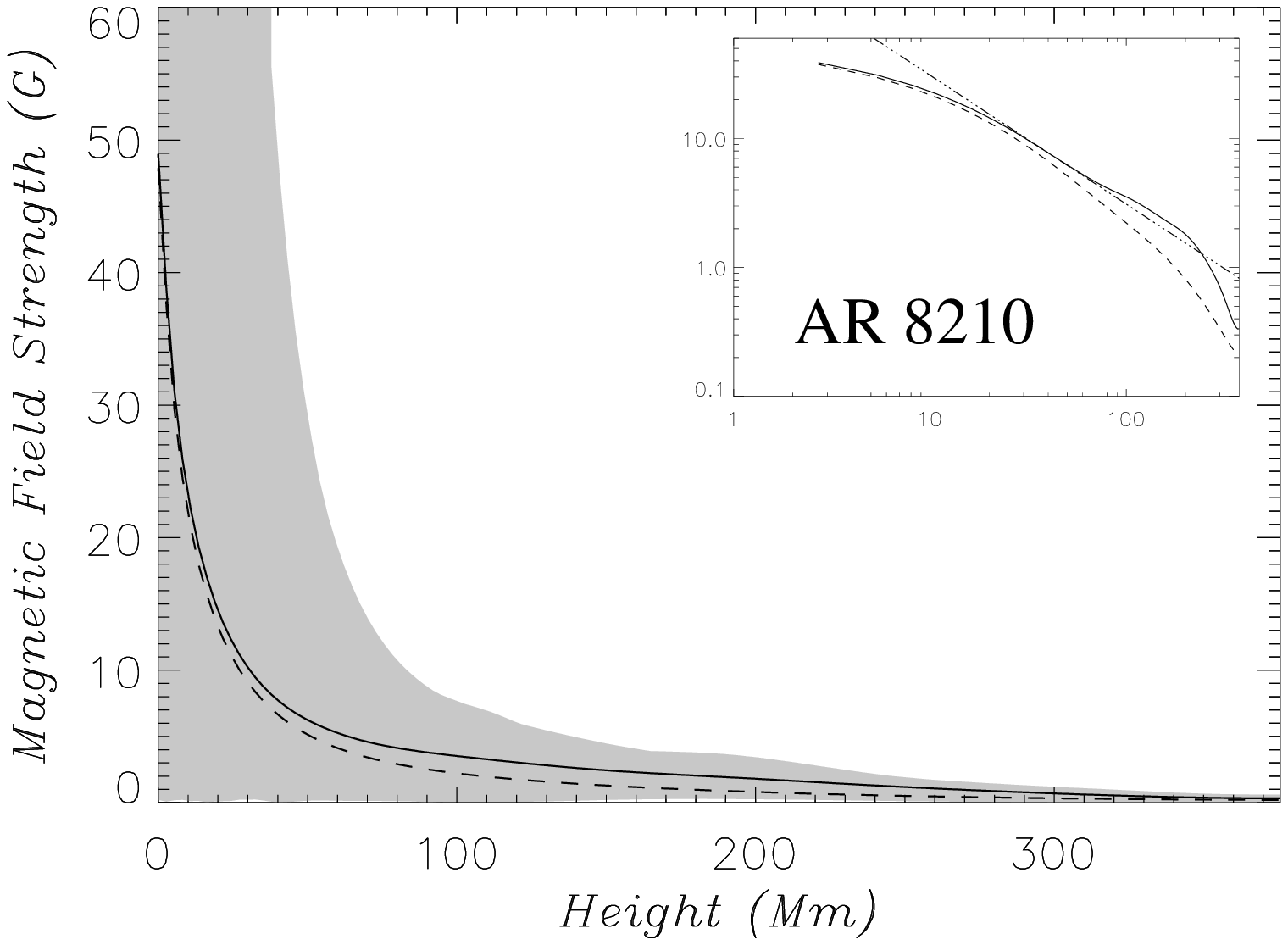}
\includegraphics[width=.49\textwidth]{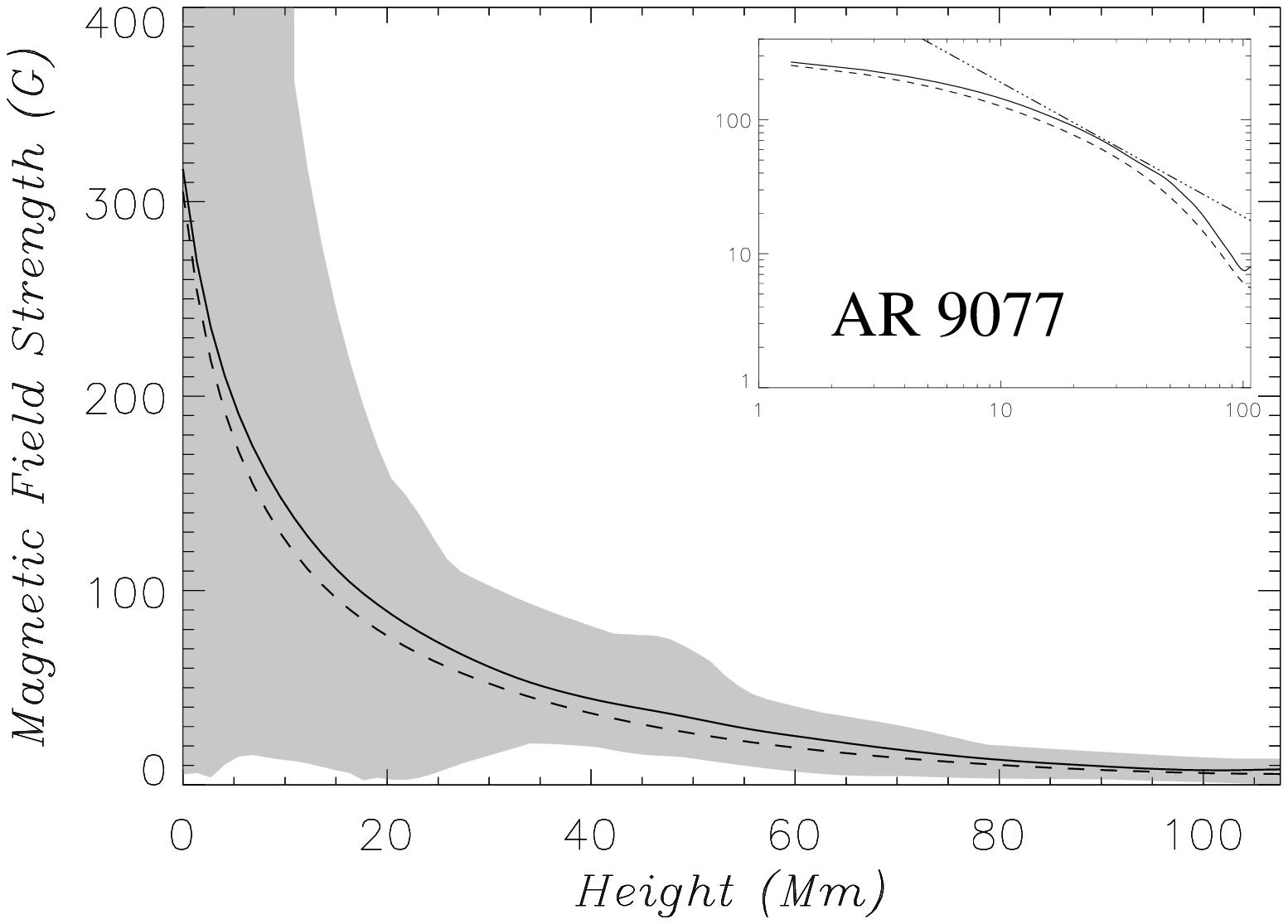}
\includegraphics[width=.49\textwidth]{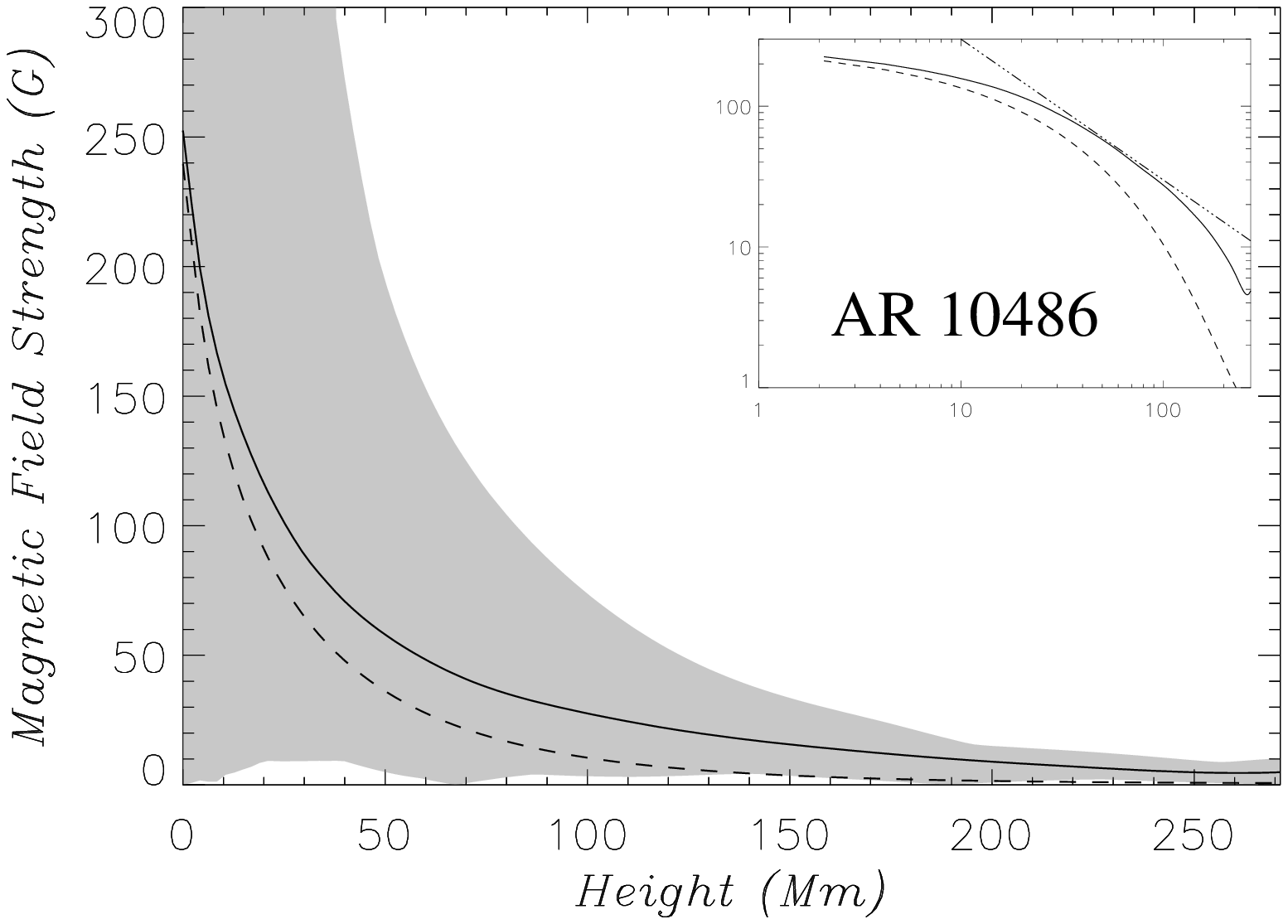}
\caption{Average magnetic field strength as a function of height for the
potential field (dashed line) and for the nonlinear force-free field (solid
line). The grey area ranges from the minimum to the maximum field strength of
the nonlinear force-free field. The log-log plots are drawn in the top-right
corners. The dot-dashed line indicates the $z^{-1}$ decay of the field.}
\label{fig:bstr}
\end{figure*}
\begin{figure*}[!ht]
\centering
\includegraphics[width=.49\textwidth]{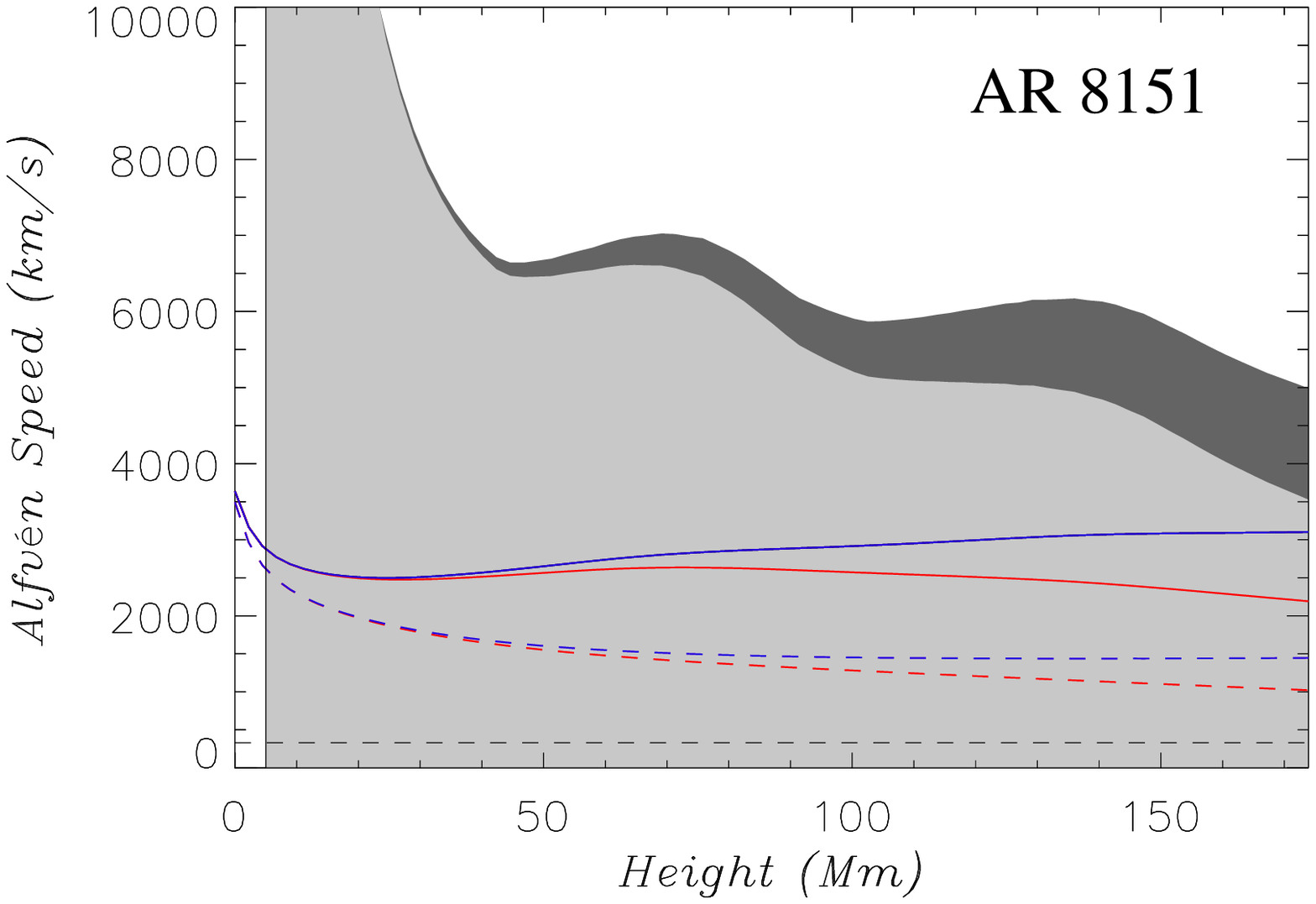}
\includegraphics[width=.49\textwidth]{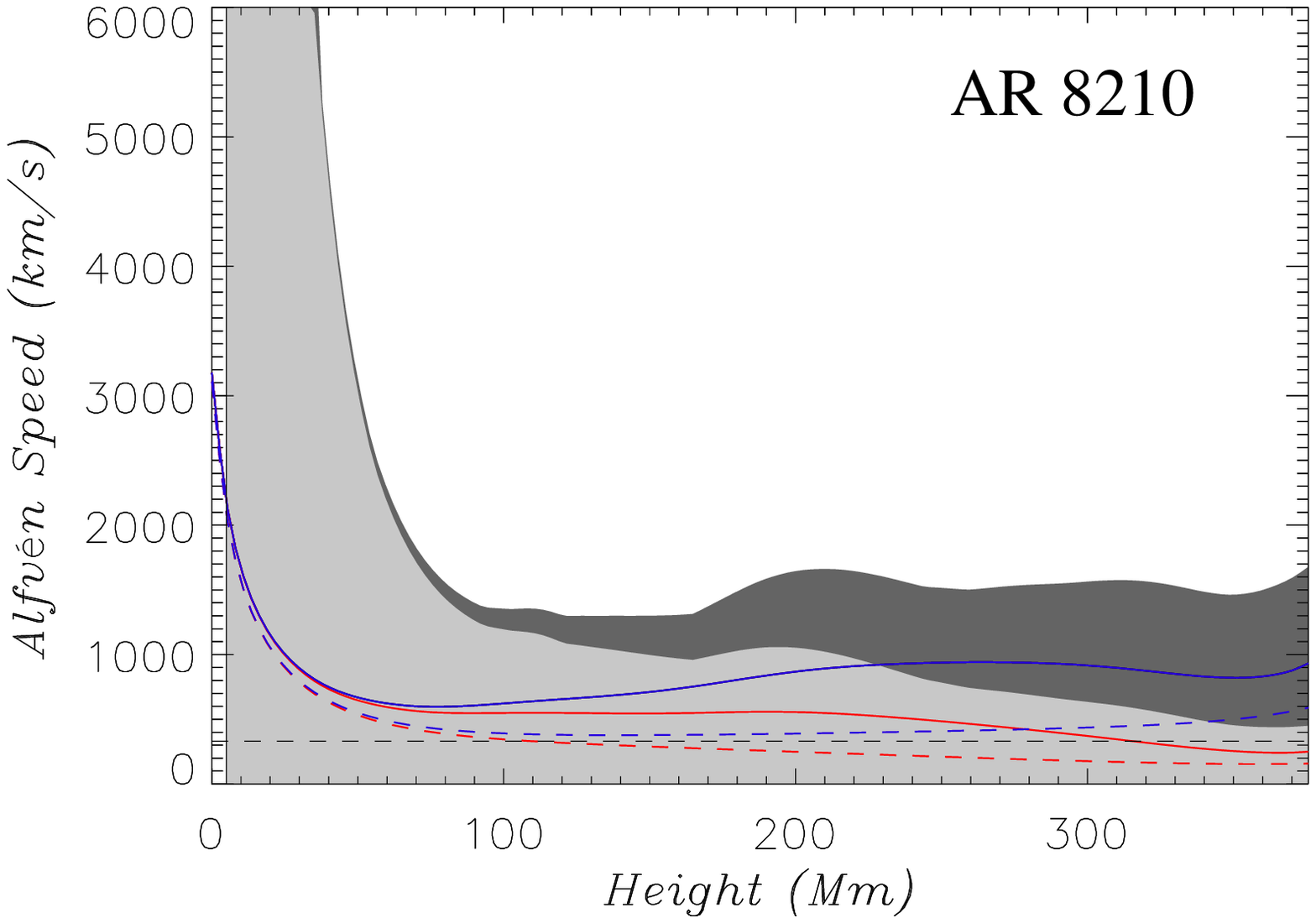}
\includegraphics[width=.49\textwidth]{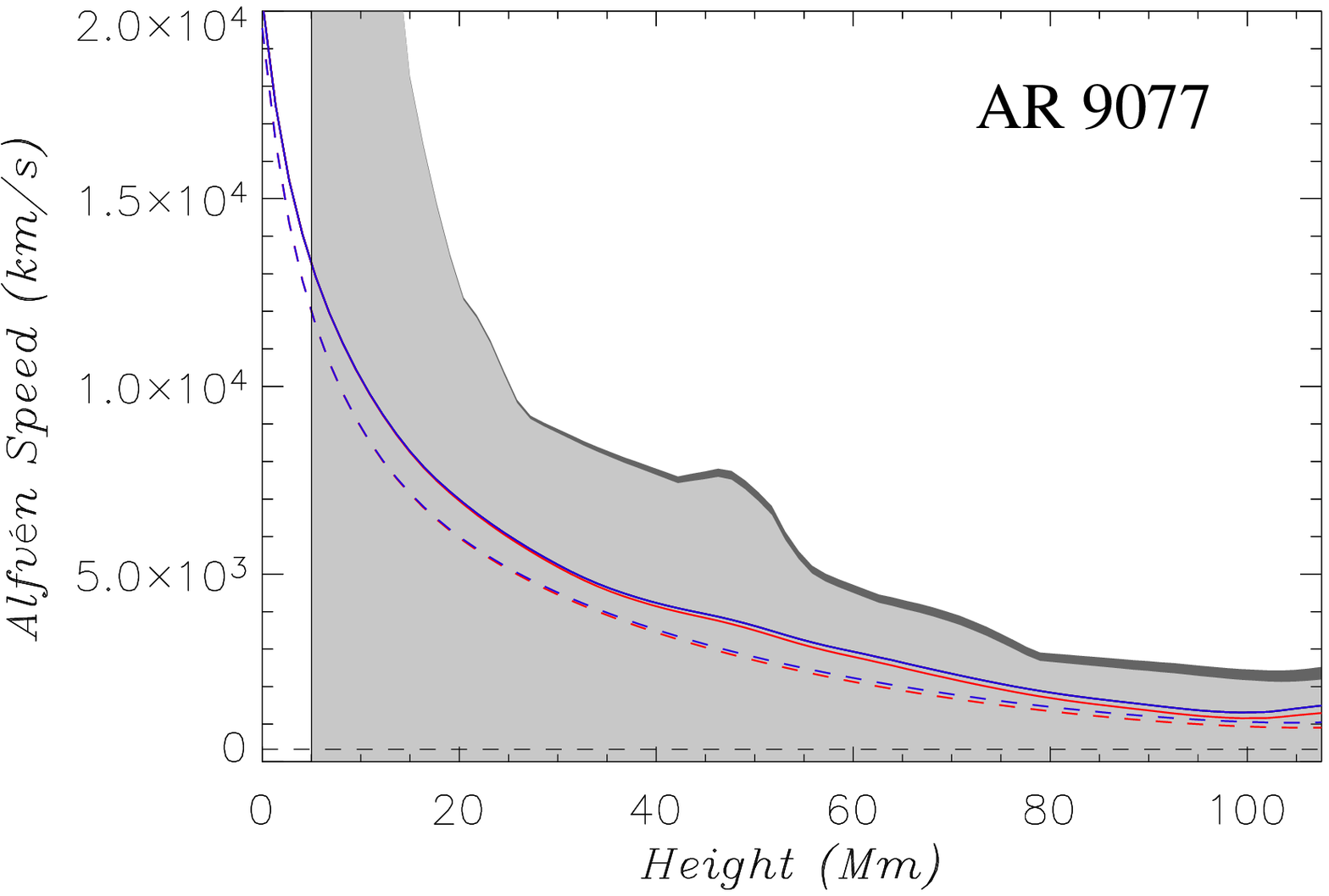}
\includegraphics[width=.49\textwidth]{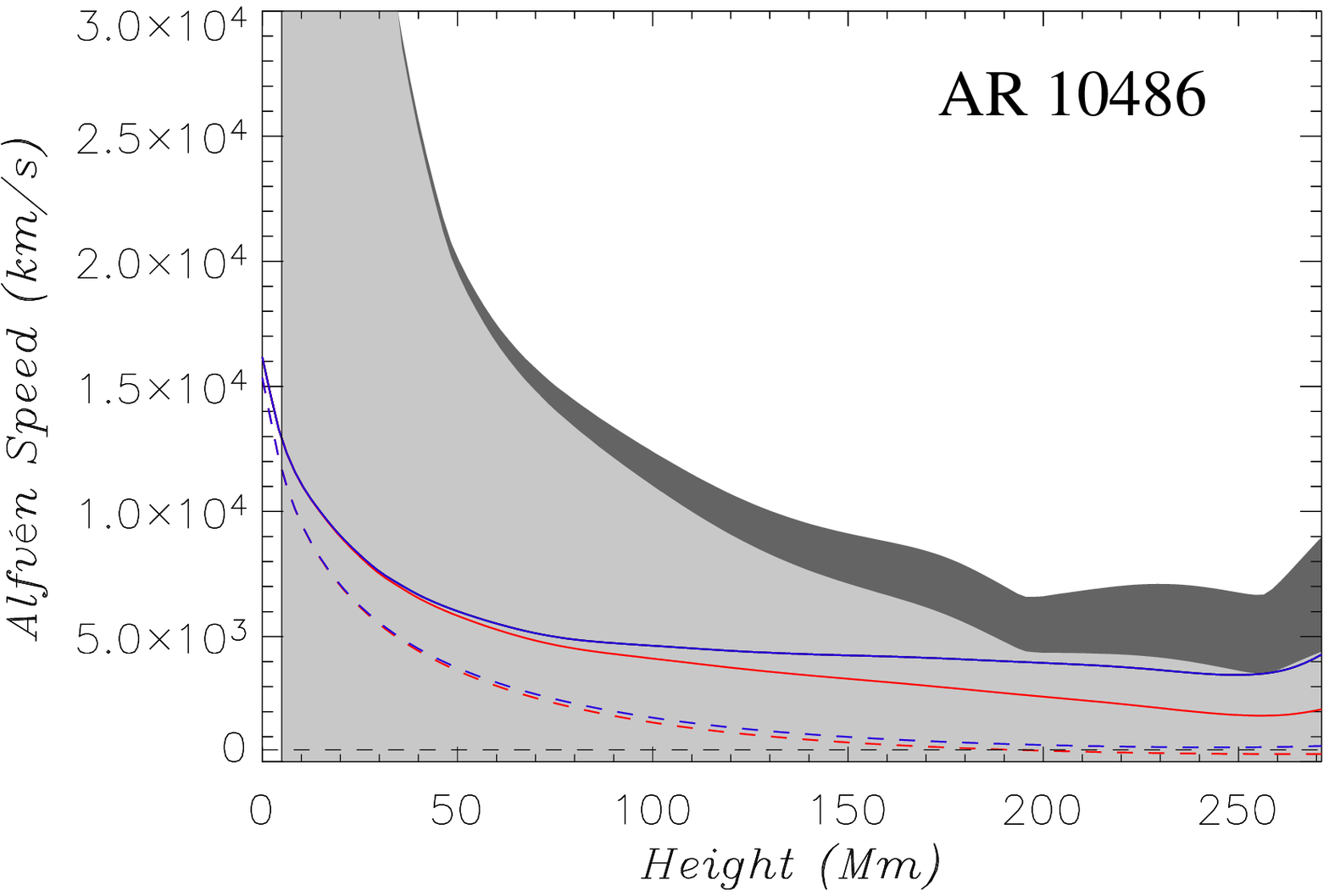}
\caption{Same as Fig.~\ref{fig:dip} for different active regions: AR 8151
(top left), AR 8210 (top right), AR 9077 (bottom left), AR 10486 (bottom right).}
\label{fig:va_iso}
\end{figure*}

\section{Active regions}
\label{sec:ar}

	\subsection{Magnetic field strength in active regions} \label{sec:bstr}

Before computing the Alfv\'en speed, we study the changes of magnetic field
strength depending on the nature of the observed active region and on the
extrapolation model. We first compute the potential and nonlinear force-free
magnetic fields associated with the four active regions described in
Section~\ref{sec:obs_ars}. Then we derive the magnetic field strength average
in an {\em xy}--plane as a function of height. In Fig.~\ref{fig:bstr}, we plot
the average field strength for the potential field (dashed line) and the
nonlinear force-free field (solid line). The grey area is the variation of
field strength for the nonlinear force-free field at a given height. 

For AR 8151 (Fig.~\ref{fig:bstr} top left), the average \nlff~field strength
ranges from 55 G on the photosphere to 9 G at 170 Mm. Assuming that the
decrease of the field strength should be exponential, there is an excess in the
values of the field strength between 40 Mm and 100 Mm where the twisted flux
bundles are located \citep[see][]{reg04, reg07}. 

For AR 8210 (Fig.~\ref{fig:bstr} top right), the average \nlff~field strength
varies from 45 G on the photosphere to small values above 300 Mm. The
\nlff~curve (solid line) follows an exponential decay as for the potential
field (dashed line). It is noticeable that the scatter of the field strength
below 50 Mm is of about two orders of magnitude. Even if the storage of
magnetic energy is focussed near the surface \citep{reg07}, the largest
departure from potential occurs between 50 Mm and 300 Mm. Below 50 Mm, the
average values of the \nlff~field strength are similar to the potential ones
but the spatial distribution is different: the free magnetic energy above the
potential field is only 2.5\% of the total magnetic energy and the energy is
stored near topological elements in the \nlff~field, as mentioned in
\cite{reg07a, reg07}.

For AR 9077 (Fig.~\ref{fig:bstr} bottom left), the total magnetic flux on the
photosphere is more important than for the above examples. The average field
strength varies from 300 G on the photosphere to 90 G at 110 Mm. The average
strength from the \nlff~field follows the same decay as for the potential
field. The active region was observed after a X5.7 flare and exhibits
post-flare loops. The variation of the field strength with height indicates
very little storage of magnetic energy in the corona due to twisted flux tubes
contrary to AR 8151.   

For AR 10486 (Fig.~\ref{fig:bstr} bottom right), the average field strength
varies from 250 G on the photosphere to 5 G at 250 Mm. It is noticeable that
the magnetic energy can be stored in the corona up to 150 Mm as seen by the
departure from the potential field (dashed curve). 

By studying the global properties of the magnetic field strength above
several active regions, we conclude that:
\begin{itemize}
\item[(i)]{the main magnetic flux concentration occurs near the photosphere up
to a height of 50 Mm;}
\item[(ii)]{a significant amount of magnetic energy can be stored in the corona
(between 50 and 150 Mm) in an active region with highly twisted flux bundles (AR
8151) or with a complex topology associated with a strong photospheric flux (AR
10486);}
\item[(iii)]{the potential and \nlff~fields have different
behaviours depending on the nature of the field: strong departure from the
potential field for active regions with highly twisted flux tubes or complex
topology associated with a high activity level, and small departure for active
regions after a strong flare or with a complex topology not related to important
flare activity.}
\end{itemize}

As pointed out by \cite{kou04}, there are few measurements of the magnetic field
strength in the solar atmosphere. The above study shows that the average
values at a given height strongly depend on the nature of the active region
(the total magnetic flux and the distribution of polarities on the photosphere)
and on the nature of the coronal magnetic field (potential or \nlff~fields). In
Section~\ref{sec:disc} we will discuss the implications for different solar
models.

Regarding the relationship between eruption and magnetic field decay, the
log-log plots (see Fig.~\ref{fig:bstr}) show several interesting features:
\begin{itemize}
\item[(i)]{the $z^{-1}$ decay line is tangent to the curve of the \nlff~field at
about 80 Mm for both active regions (AR 8151 and AR 10486) associated with an
eruption causing a large-scale magnetic field reorganisation, whilst for a
post-eruptive active region (AR 9077) and the one associated with confined
flares (AR 8210), the $z^{-1}$ line is at about 20 Mm;}
\item[(ii)]{the weak slope of the log-log plots below the $z^{-1}$ line
indicates the strong influence of the current systems as in \cite{van78}.}
\end{itemize}

The implications for eruption models are developed in Section~\ref{sec:disc}.

	\subsection{Alfv\'en speed in solar active regions} \label{sec:va_ar}

We apply the above models to the active regions described in
Section~\ref{sec:obs_ars}. To avoid confusion, we only report here the
Alfv\'en speed values for the constant gravity model. In Fig.~\ref{fig:va_iso},
we plot the average Alfv\'en speed values in the {\em xy}--plane as a function
of height derived from the potential field (dashed curves) and \nlff~field
(solid curves) for the different active regions. First, the average Alfv\'en
speed at the base of the corona varies from 2--3000 \kms~for AR 8151 and AR 8210
to 13000 \kms~for AR 9077 and AR 10486. It is strongly related to the total
magnetic flux of the active region on the photosphere. Above the coronal base,
the variation of the Alfv\'en speed with height is greatly influenced by the
nature of the field: the topology, the geometry of field lines and the effects
of currents. 

\begin{figure}[!h]
\centering
\includegraphics[width=1.\linewidth]{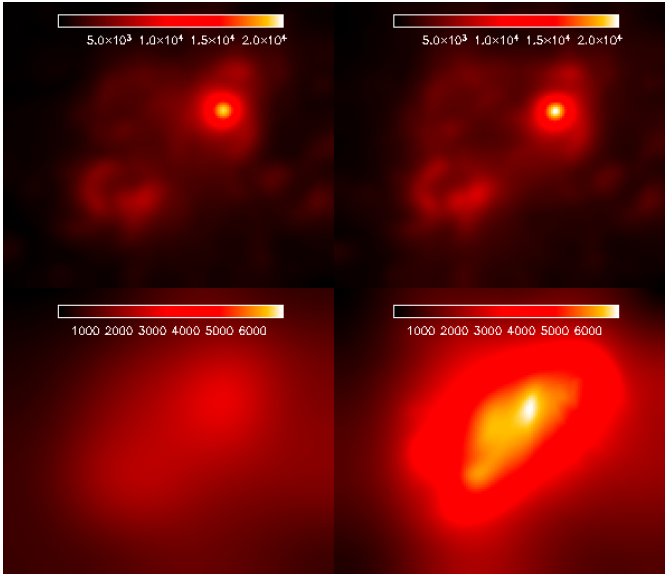}
\caption{Alfv\'en speed dynamic range in AR 8151 for the potential field (left
column) and the nonlinear force-free field (right column) at two different
heights, 10 Mm (top) and 60 Mm (bottom).}
\label{fig:va_ar8151}
\end{figure}

For AR 8151, the Alfv\'en speeds in the \nlff~field can reach up to 52000
\kms~near the base of the corona (for an average value of 3000 \kms). Then the
maximum Alfv\'en speed values decrease to 6000 \kms~at 50 Mm to finally reach
the value of 5000 \kms~above 150 Mm. The evolution of the average values is
almost constant at 3000 \kms. Nevertheless there is a minimum of Alfv\'en speed
of 2500 \kms~at 25 Mm. This result is consistent with \cite{war05}. According to
\citet{reg07}, the minimum is below the height where the twisted flux tubes are
located and the magnetic energy is deposited in this active region. The
evolution of the average speeds in the potential field (dashed curve) is
strictly decreasing from 2500 \kms~to 1500 \kms. Note that according to
Section~\ref{sec:model_param}, the average speeds should increase above 170 Mm
when the density dominates. Both potential and \nlff~average speed curves are
above the $\beta < 1$ limit (dashed line). Even if the average values of the
Alfv\'en speed for the potential and \nlff~fields are not far from each other in
the low corona, there is a factor of about 2 difference above 80 Mm. In
Fig.~\ref{fig:va_ar8151}, we draw the dynamic range of Alfv\'en speed at two
different heights (top: 10 Mm, bottom: 60 Mm) for both potential (left) and
\nlff~(right) fields. At both heights and for both models, the greater is the
magnetic field strength, the larger is the Alfv\'en speed. Even if the dynamic
range is similar for the potential and \nlff~fields at 10 Mm, the Alfv\'en speed
varies by a factor of 2 at 60 Mm due to the presence of twisted flux bundles
along the polarity inversion line. The average Alfv\'en speed decreases
more rapidly for the varying gravity model: at a height of 150 Mm, the average
Alfv\'en speed is 2400 \kms~for a varying gravity and 3000 \kms~for a constant
gravity.

\begin{figure}[!h]
\centering
\includegraphics[width=1.\linewidth]{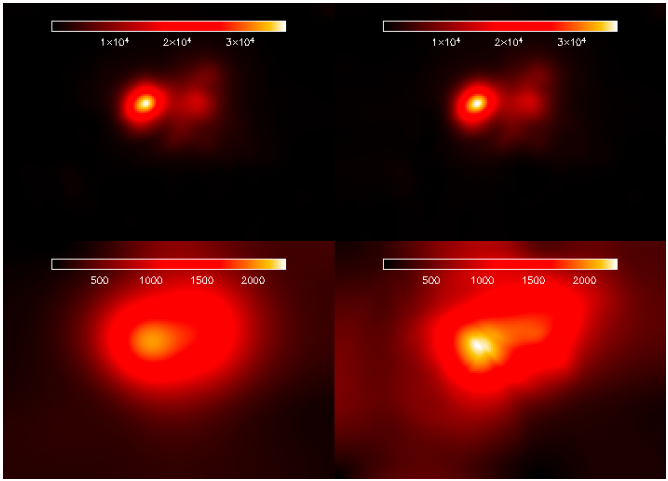}
\caption{Alfv\'en speed dynamic range in AR 8210 for the potential field (left
column) and the nonlinear force-free field (right column) at two different
heights, 10 Mm (top) and 60 Mm (bottom).}
\label{fig:va_ar8210}
\end{figure}

For AR 8210, the maximum Alfv\'en speed at the base of the corona is about 70000
\kms. Even if the total magnetic flux is comparable to the flux of AR 8151, the
maximum field strength is larger in the main sunspot of AR 8210 \citep{reg06}
which explains the larger value of the maximum Alfv\'en speed. The average speed
for the \nlff~field decreases rapidly from the base of the corona to about 70 Mm
from 2200 \kms~to 600 \kms. The fast decrease is explained by the nature of the
magnetic configuration: the confined magnetic field in the active region, and
the storage of magnetic energy in the low corona \citep[see][]{reg06}. The
average speed for the potential field follows the same fast decrease below 70 Mm
before reaching a minimum at about 150 Mm. Above this height, the average speed
slowly increases when the density dominates the field strength. At 100 Mm, the
average speed for the \nlff~field (potential resp.) is about 650 \kms~(400
\kms~resp.) which is indeed of factor of 1.5 more for the \nlff~field. The
dynamic range of the Alfv\'en speed is depicted in Fig.~\ref{fig:va_ar8210}.
There are very few changes between the dynamic range images at a given height.
We then suggest that the complex topology of AR 8210 does not significantly
influence the Alfv\'en speed distribution. This conclusion is also supported by
the fact that the magnetic energy is deposited near the photospheric boundary
and not in the corona. Since the average Alfv\'en speed decreases faster for
the varying gravity model, the Alfv\'en speed becomes less than the
coronal sound speed at about 300 Mm, whilst the Alfv\'en speed stays above the
plasma $\beta = 1$ limit for the constant gravity model.

\begin{figure}[!h]
\centering
\includegraphics[width=1.\linewidth]{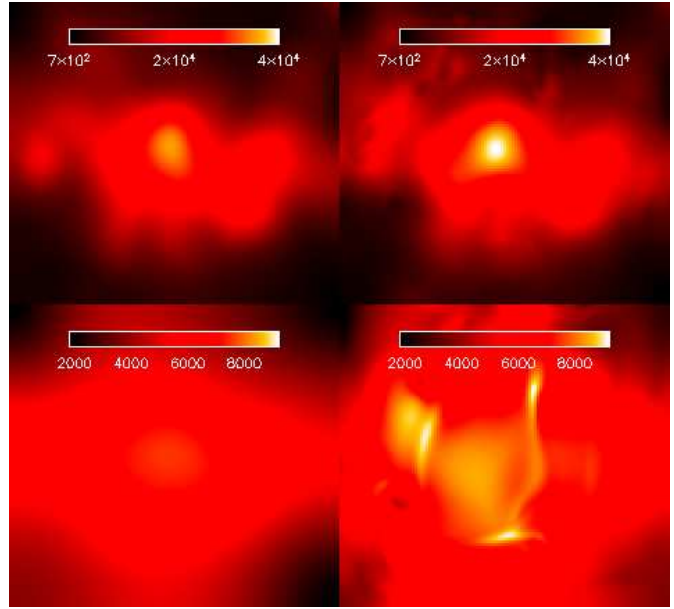}
\caption{Alfv\'en speed dynamic range in AR 9077 for the potential field (left
column) and the nonlinear force-free field (right column) at two different
heights, 10 Mm (top) and 60 Mm (bottom).}
\label{fig:va_ar9077}
\end{figure}

\begin{figure}[!h]
\centering
\includegraphics[width=1.\linewidth]{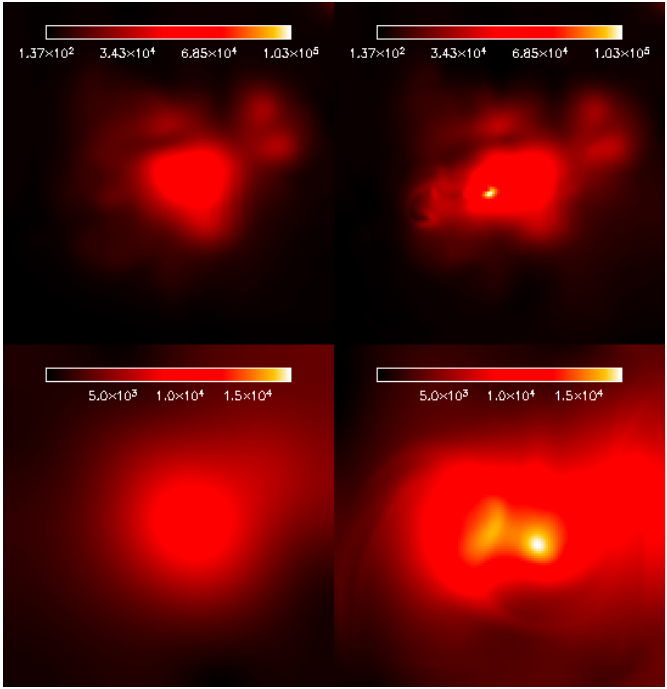}
\caption{Alfv\'en speed dynamic range in AR 10486 for the potential field (left
column) and the nonlinear force-free field (right column) at two different
heights, 10 Mm (top) and 60 Mm (bottom).}
\label{fig:va_ar0486}
\end{figure}
For AR 9077, the total magnetic flux included in the field-of-view is larger
than for the previous two examples. Therefore, even if the maximum Alfv\'en
speed at the base of the corona is moderate ($\sim$ 46000 \kms), the average
speed is relatively large (13000 \kms). Both average curves for the potential
and \nlff~field are rapidly and strictly decreasing towards 1500 \kms. Contrary
to the previous example, there is no evidence of storage of energy in the middle
corona. The strong flare and the CME associated with this active region are
related to a complex topology and the mass supply is provided by a low filament
on the side of the active region \citep[e.g.,][]{zha02}. The reconstructed
active region is performed from a magnetogram recorded after the flare. As
suggested by Fig.~\ref{fig:va_ar9077}, the dynamic ranges are quite different
from one model to another, whatever the altitude we look at, even if the
Alfv\'en speed values are comparable. Therefore the post-eruptive configuration
is strongly influenced by the electric currents flowing along field lines
despite a magnetic geometry which looks similar to potential. For both
gravity models, the average Alfv\'en speed curves are similar indicating that,
during this relaxation phase, gravity does not greatly influence the magnetic
field and the plasma.
		
For AR 10486, the maximum Alfv\'en speed at the base of the corona is estimated
to be 130000 \kms, which is about one-third of the speed of light. The
average speed for both potential and \nlff~fields is about 13000 \kms. The
average speed for the \nlff~field is decreasing rapidly from 13000 \kms~at 5 Mm
to 6000 \kms~at 50 Mm and then reaches an almost constant value of 5000 \kms.
In the view of the previous examples, we interpret this fact as a consequence
of the storage of magnetic energy in the corona at a height of about 50 Mm
either by twisted flux tubes or by sheared arcades. Following \citet{reg04f},
the existence of sheared arcades in a complex topology is confirmed. The
effects of the transverse field components and so of the electric currents are
shown in Fig.~\ref{fig:va_ar0486}. The location of strong magnetic field
strength is different from one model to another showing the complex distribution
of the magnetic field on the photosphere \cite[e.g.,][]{reg04f, man06}. The
Alfv\'en speed for the constant gravity model is almost twice the value for the
varying gravitational field high in the corona. 
\begin{figure*}[!ht]
\centering
\includegraphics[width=.49\linewidth]{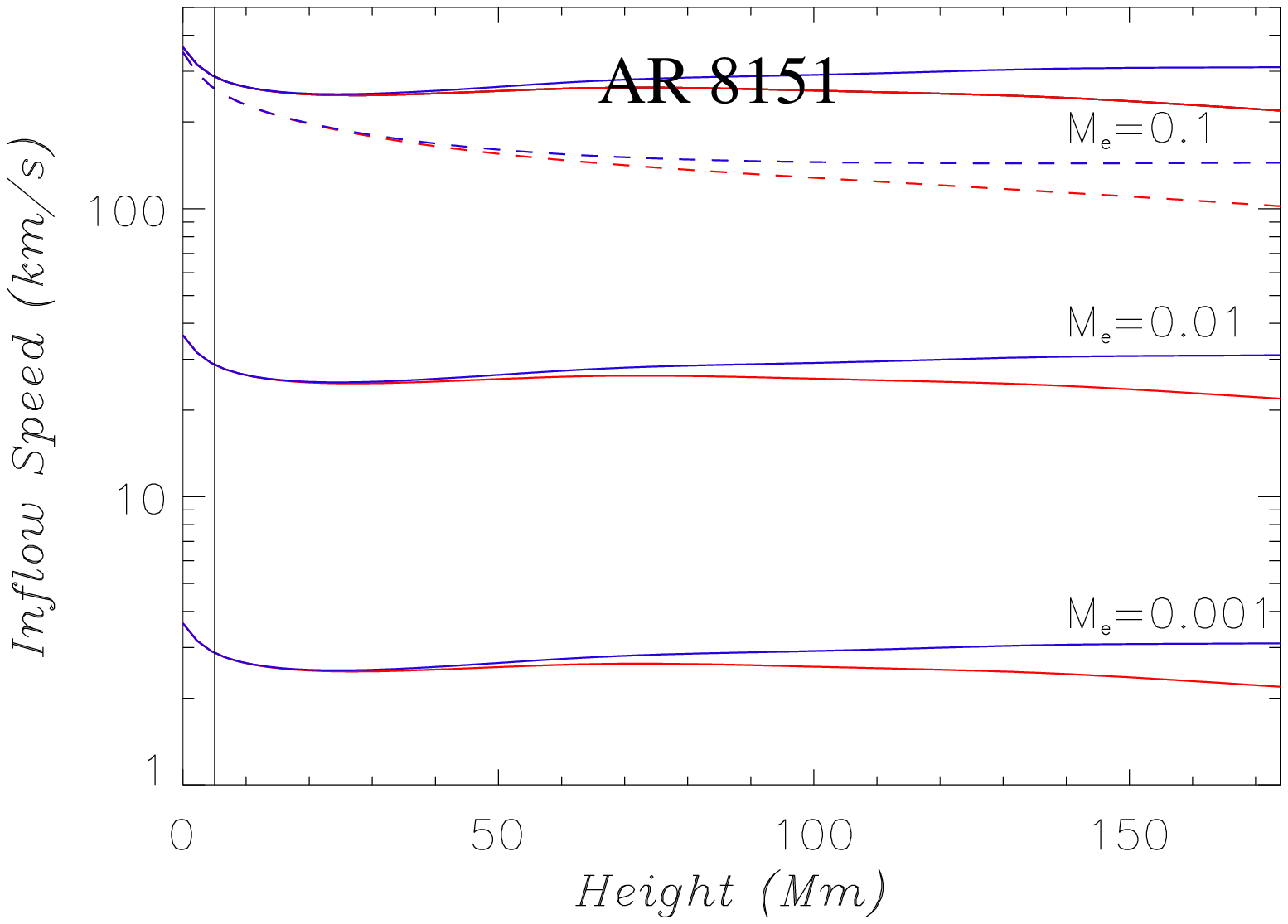}
\includegraphics[width=.49\linewidth]{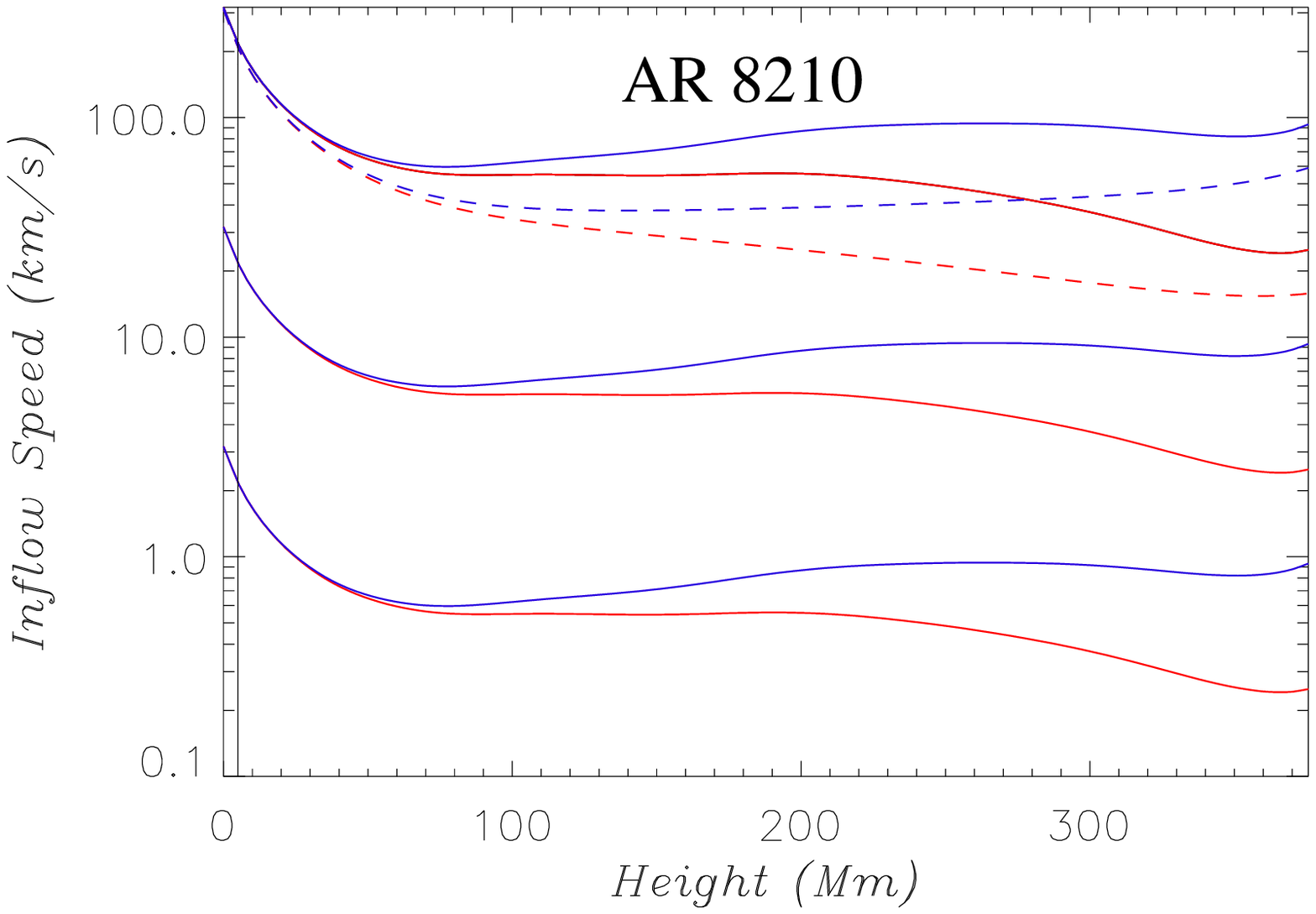}
\includegraphics[width=.49\linewidth]{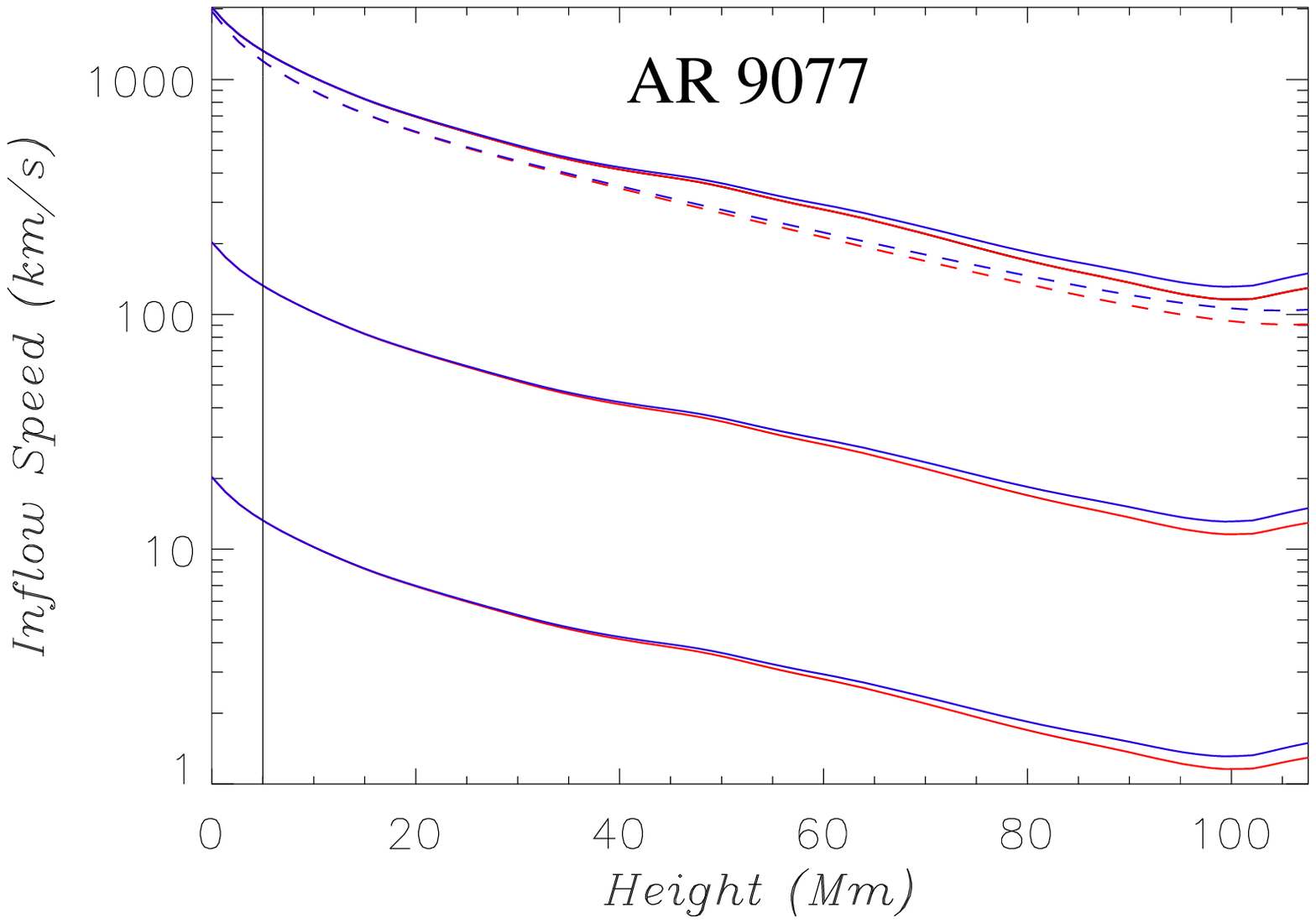}
\includegraphics[width=.49\linewidth]{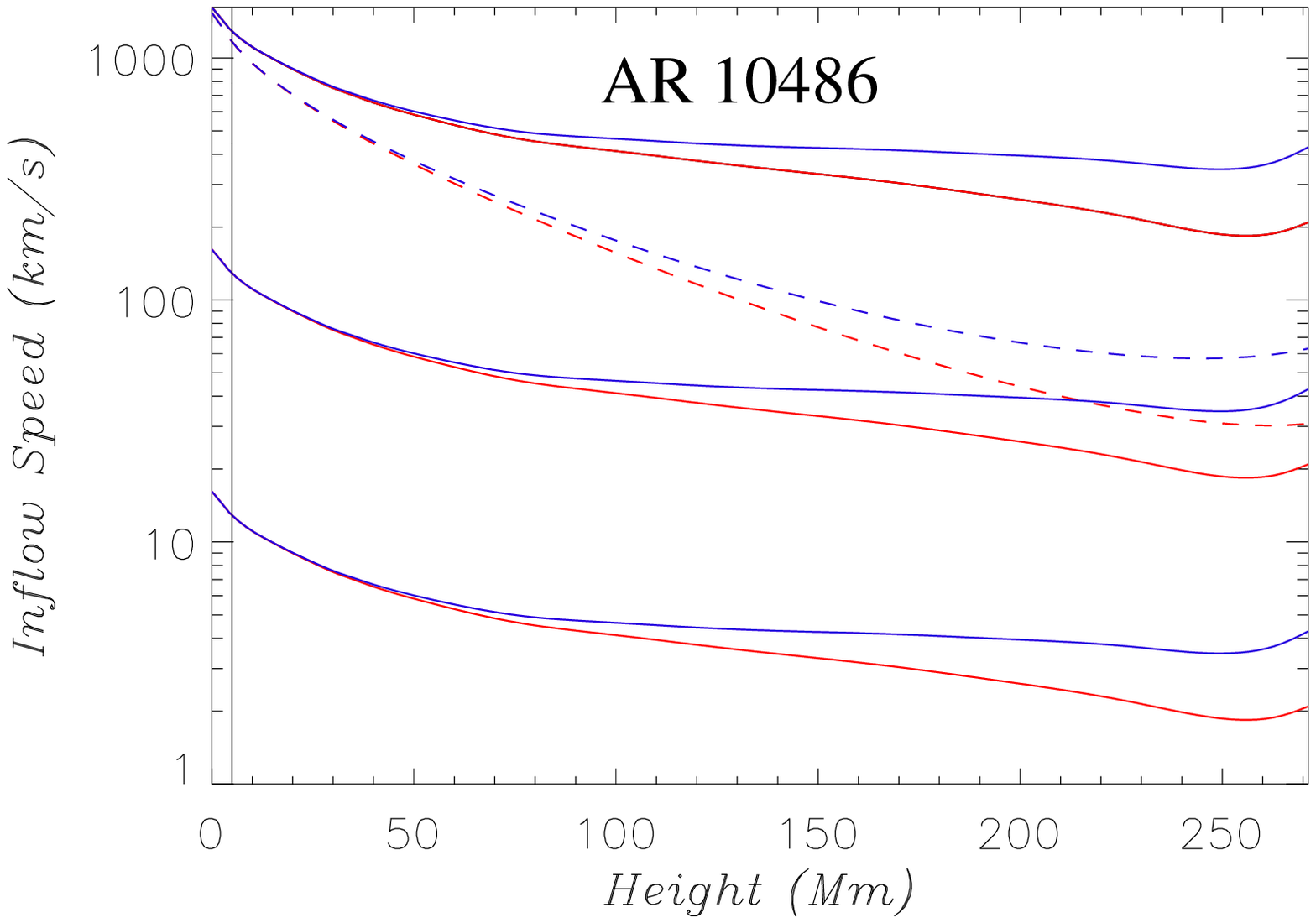}
\caption{Average inflow speeds as a function of height for the nonlinear
force-free field (solid line) and potential field (dashed line), with a
reconnection rate $M_e$ varying from 10$^{-3}$ to 10$^{-1}$. Red (resp. blue)
curves are for the varying (resp. constant) gravity model.}
\label{fig:inflow}
\end{figure*}

\section{Discussions and Conclusions}
\label{sec:disc}

\begin{figure*}
\centering
\includegraphics[width=.49\linewidth]{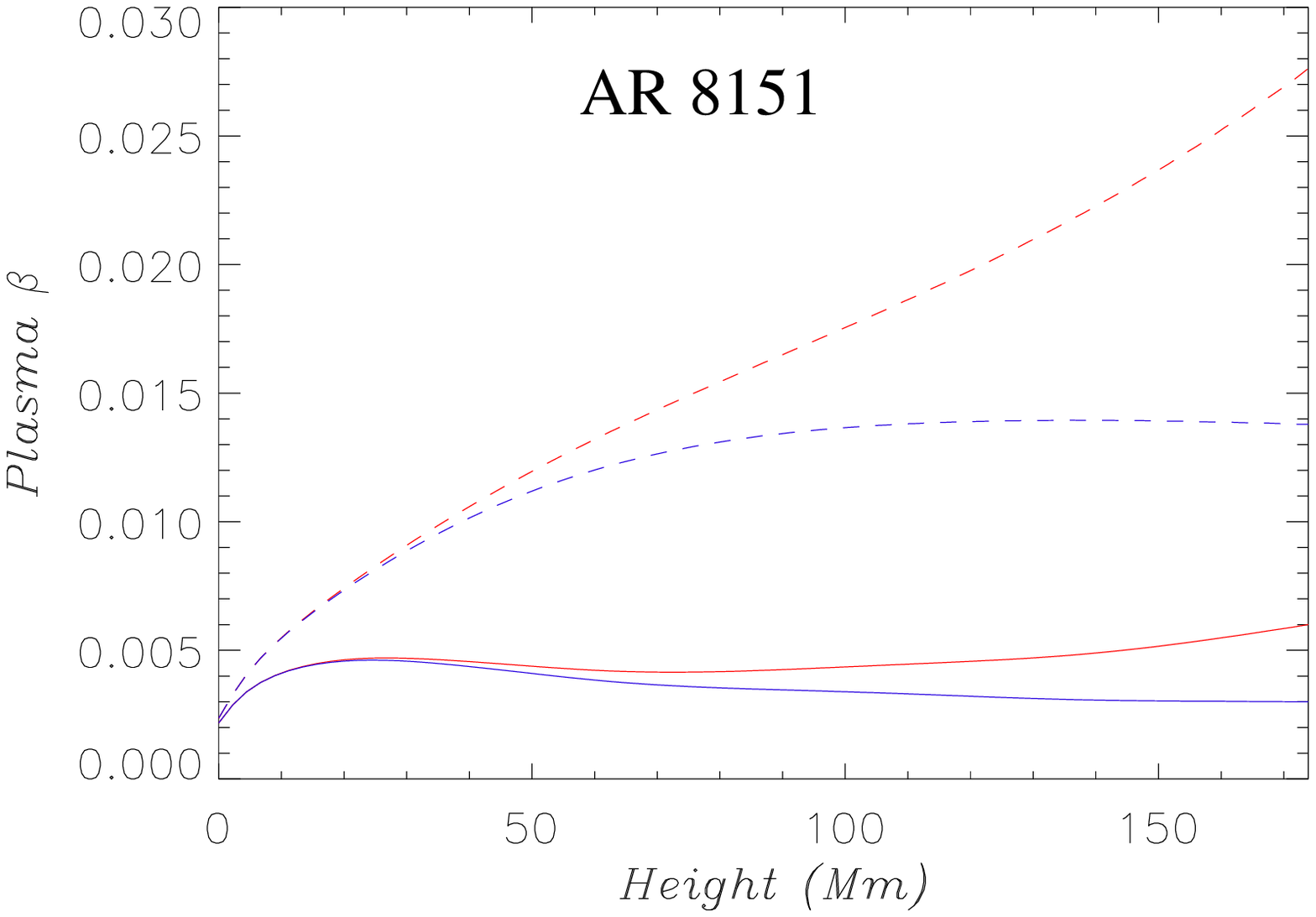}
\includegraphics[width=.49\linewidth]{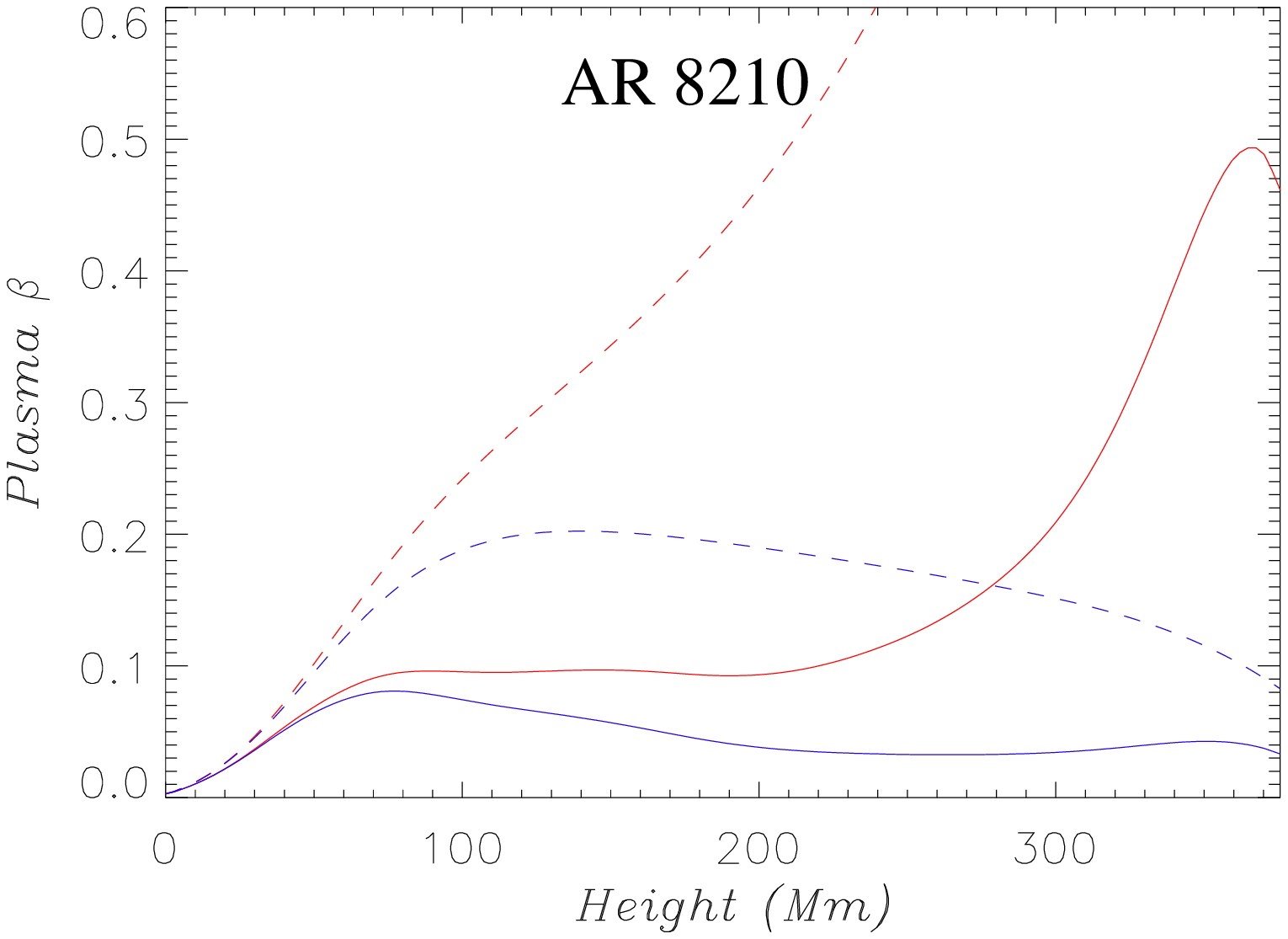}
\includegraphics[width=.49\linewidth]{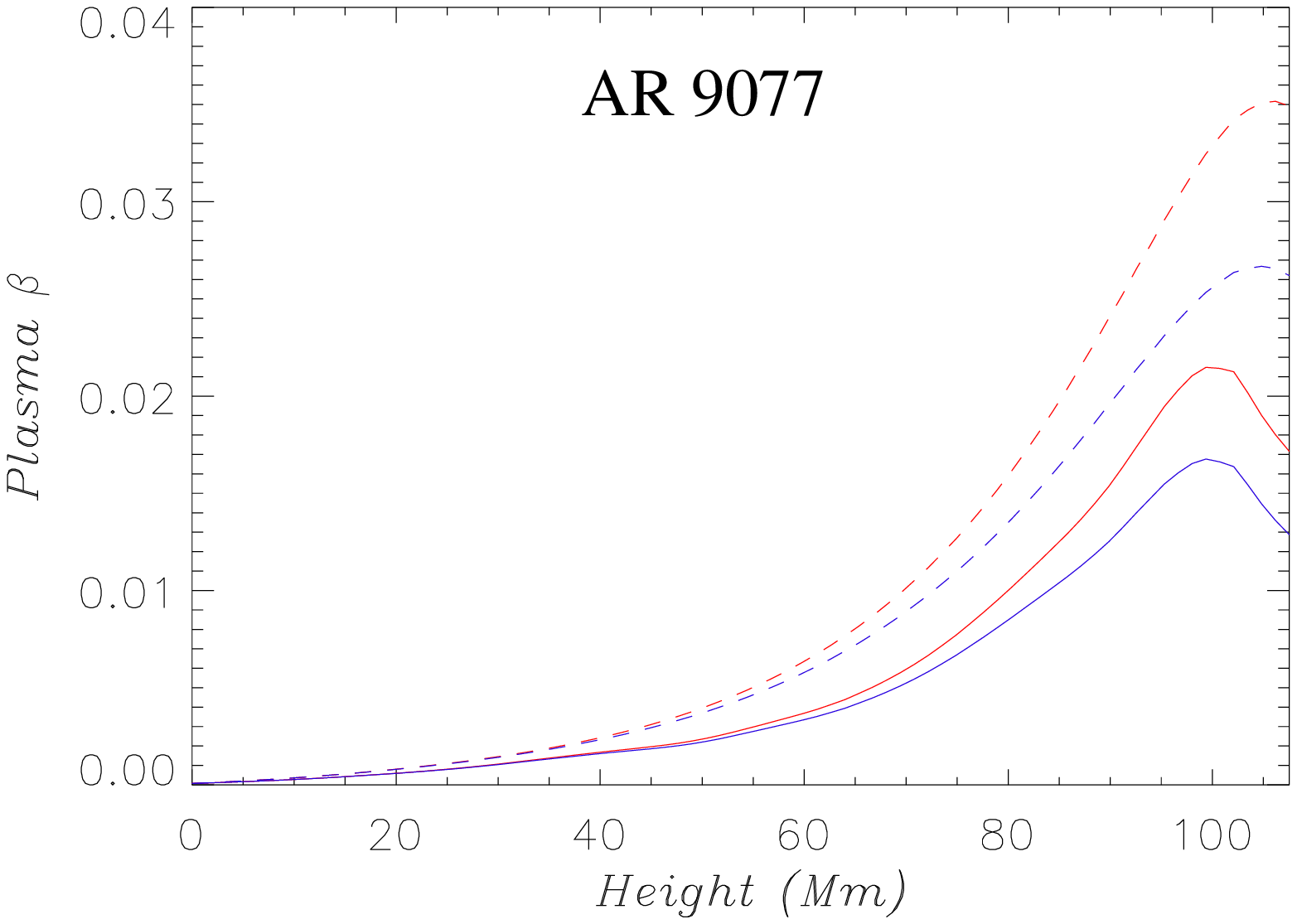}
\includegraphics[width=.49\linewidth]{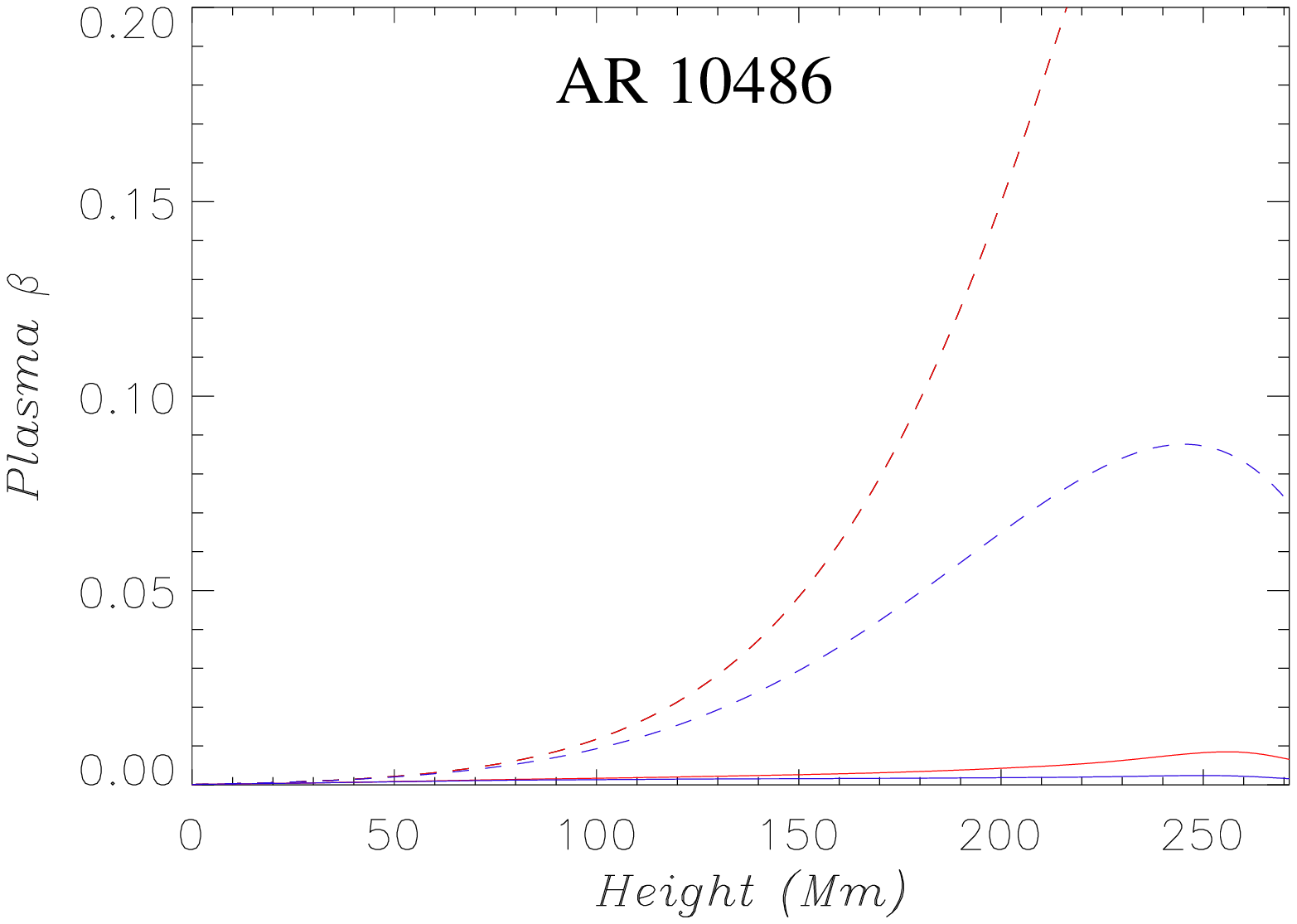}
\caption{Plasma $\beta$ for the nonlinear force-free field (solid line) and the
potential field (dashed line) as a function of height for four active regions.
Red (resp. blue) curves are for the varying (resp. constant) gravity model.}
\label{fig:beta}
\end{figure*}
In this paper, we address two questions related to the global properties of the
magnetic field and the plasma in the corona:
\begin{itemize}
\item[1-]{How does the magnetic field strength and Alfv\'en speed vary with height
in the solar corona?}
\item[2-]{What can we learn from the estimate of the coronal Alfv\'en speed
for some key solar physics problems (coronal heating, magnetic reconnection,
wave propagation, initiation and development of CMEs)? }
\end{itemize}

In order to tackle the above questions, we have developed a method to determine
the plasma properties in a magnetic field above active regions. We assume that
an active region can be modelled in two steps: a nonlinear force-free
equilibrium for the magnetic field, and a hydrostatic equilibrium between
pressure and gravity forces describing the plasma properties. To solve the
latter equation, the corona is assumed here to be isothermal with either a
constant gravity or a gravitational field varying with height.

In Sections~\ref{sec:bdip} and~\ref{sec:bstr}, we investigated the
evolution of the average field strength with height. Interestingly, the log-log
plots showed that the decay of the magnetic field with height does not follow a
power law, even for a potential field. This fact suggests that the decay of the
field strength is strongly influenced by the geometry of the field (e.g.,
toroidal geometry for the bipolar field), by the distribution of electric
currents and their typical height of influence, and by the complexity of the
magnetic distribution on the photosphere for active regions. This is a
well-known property of the magnetic field. Based on the data reported by
\cite{pol75}, \cite{van78}  plotted the measured magnetic field strength as a
function of height for three different active regions and assuming current-free
magnetic configurations. They obtained similar curves to those shown in this
paper in which the log-log plot can be fitted by a parabola. In addition, they
fitted the distribution of field strength by two power laws of index -1 and -3
suggested by their theoretical work: -1 for a line current along the inversion
line of the magnetic field and -3 for a dipolar potential field decay.
\cite{van78}  concluded that (i) the electric currents strongly influence the
magnetic configuration when the power law index is less than -1, (ii) above a
certain height the magnetic field decays rapidly (probably with a power law
index of -3). More importantly, the authors concluded that, if the magnetic
field below the $z^{-1}$ decay is decaying with a power law index less than -1,
then the configuration is stable. This condition can be compared with the
conditions required to trigger a kink instability \citep[e.g.,][]{hoo81} or a
torus instability \citep{tor06}. Recently,  \cite{liu08} investigated the
evolution of the magnetic field strength with height and compared with the
nature of the eruptive phenomenon. He derived the magnetic field from a
large-scale potential field extrapolation based on the PFSS model (Potential
Field Source Surface, e.g., \citeauthor{sch69} \citeyear{sch69}). For heights
between 42 Mm and 140 Mm, the power law index is about -1.5 whatever the onset
mechanism. Combined with our study, this result suggests that the behaviour of
the magnetic field in the  low atmosphere plays the main role in triggering
flares and CMEs, and also that electric currents flowing along field lines as in
the \nlff~model affect strongly the power law index and therefore the stability
of the magnetic configuration.
\begin{figure*}
\centering
\includegraphics[width=.49\linewidth]{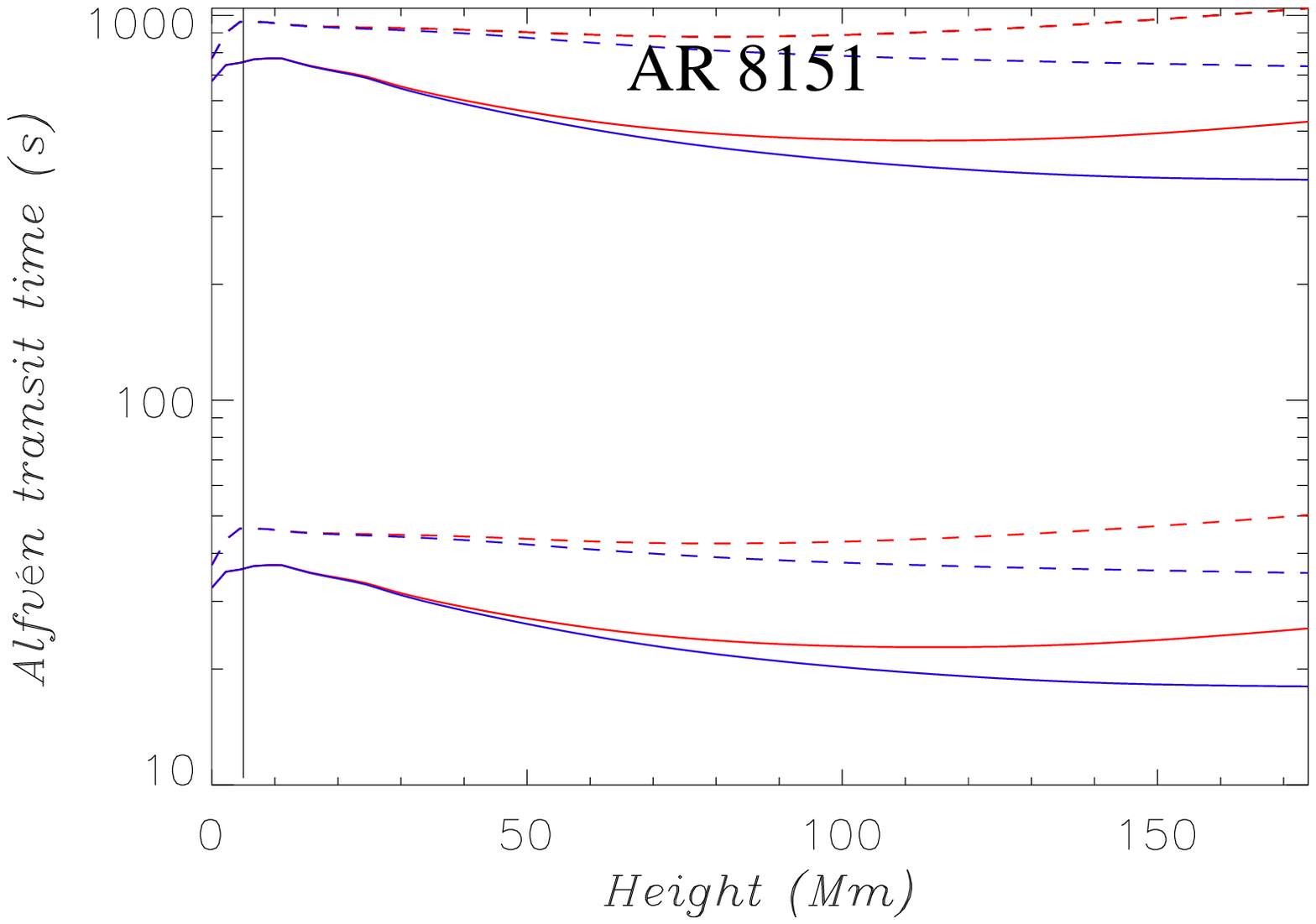}
\includegraphics[width=.49\linewidth]{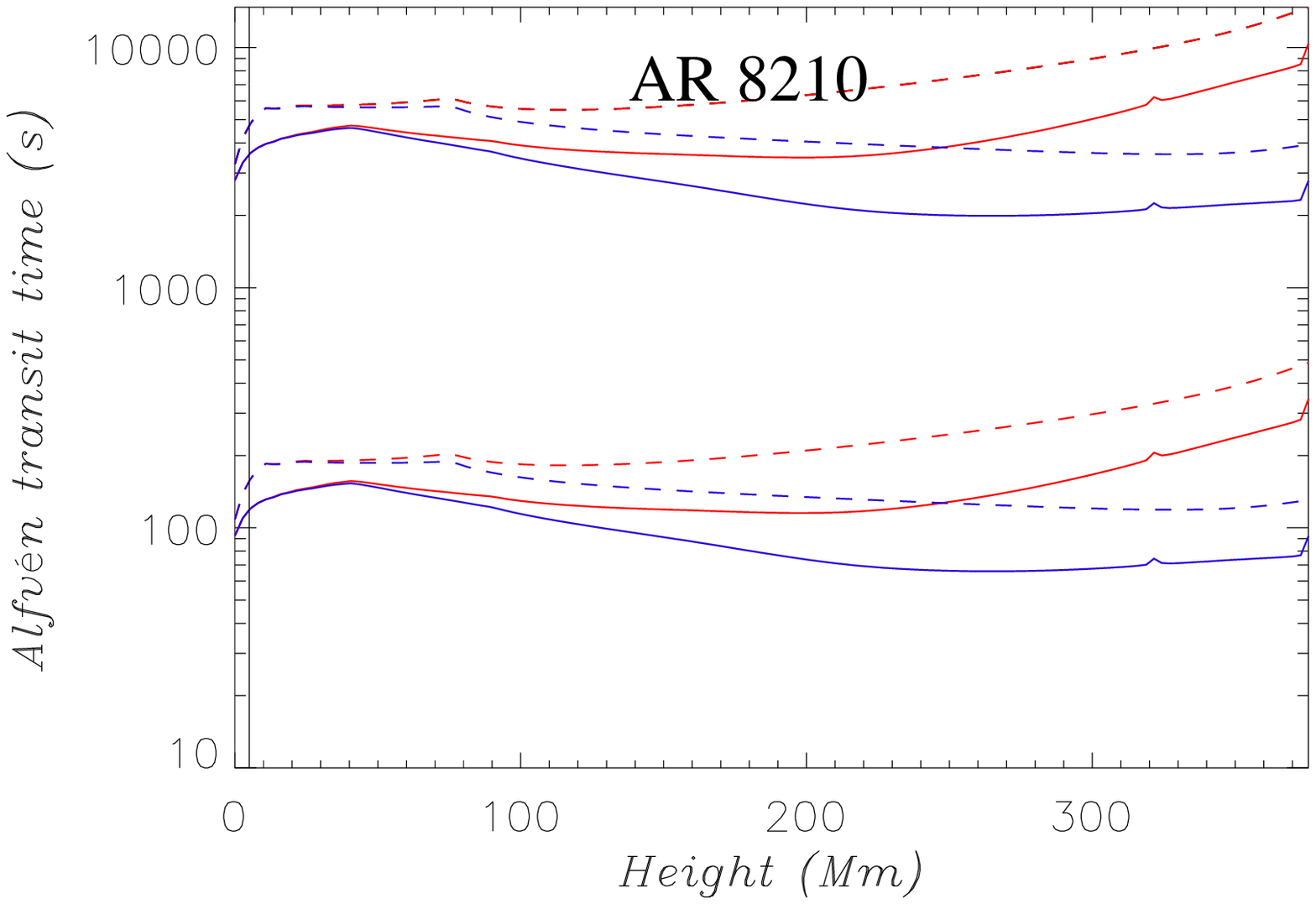}
\includegraphics[width=.49\linewidth]{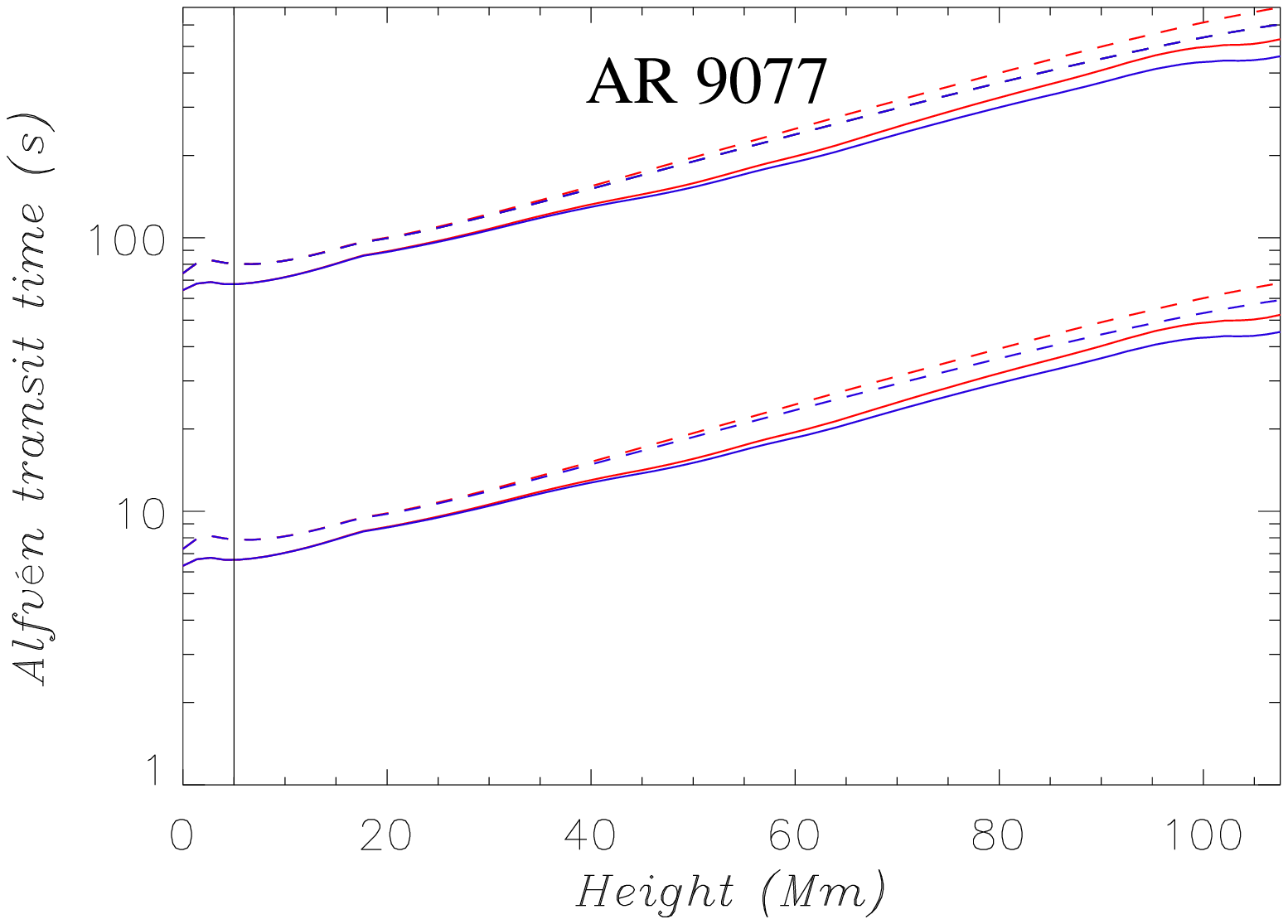}
\includegraphics[width=.49\linewidth]{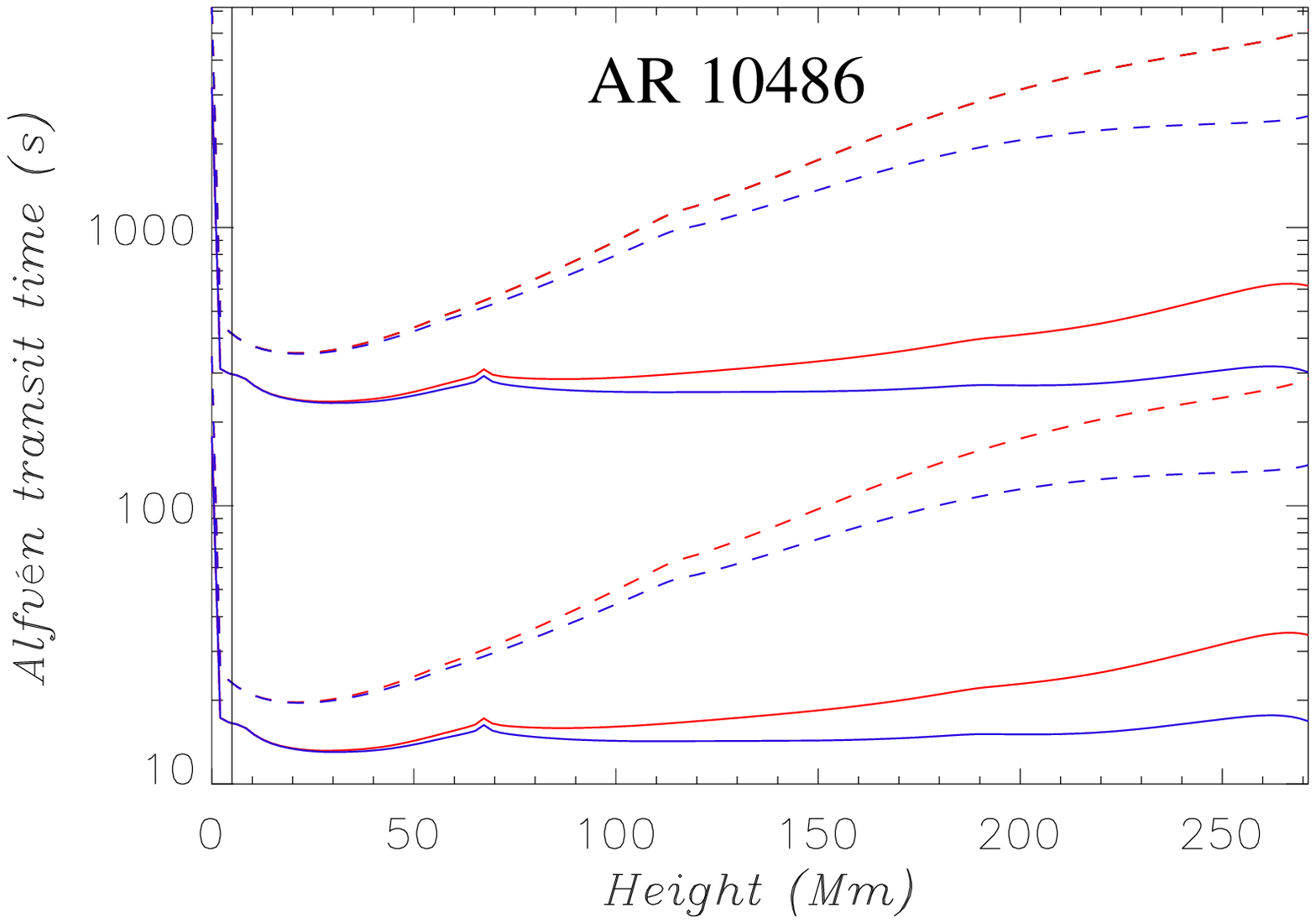}
\caption{Average Alfv\'en transit time for the potential (dashed line) and 
nonlinear force-free (solid line) fields using as typical scale the pressure
scale-height $H$ (bottom curves) or the length of a semi-circular loop with a
diameter of the scale of the computational box (top curves). Red (resp. blue)
curves are for the varying (resp. constant) gravity model.}
\label{fig:tau}
\end{figure*}

	\subsection{Reconnection rate}
	
As mentioned in Section~\ref{sec:intro}, the Alfv\'en speed is an important
parameter in understanding the magnetic reconnection processes. The reconnection
is characterised by the reconnection rate (or external Alfv\'en Mach number) as
follows:
\begin{equation}
M_e = \frac{v_i}{v_A}, 
\end{equation}  
where $v_i$ is the inflow speed. This definition does not depend on the nature
of the diffusion region where the magnetic reconnection occurs. To distinguish
between different regimes of reconnection, the magnetic Reynolds number is
introduced as follows:
\begin{equation}
R_{me} = \frac{\eta}{L_e v_A},
\end{equation}
where $\eta$ is the magnetic diffusivity, and $L_e$ is a typical length of the
diffusion region \citep[e.g.,][]{pri00}. If $R_{me}^{-1} < M_e <
R_{me}^{-1/2}$, we have a slow reconnection regime of the Sweet-Parker type. If
$M_e > R_{me}^{-1/2}$, the reconnection regime is fast such as in the
\cite{pet64} model. For a typical astrophysical plasma, the reconnection rate in
the fast regime is in the range 0.01 to 0.1. The square of the reconnection
rate is also the ratio of kinetic energy to magnetic energy of the inflow
region assuming that the diffusion region is small enough to have the same
field strength in the inflow and outflow regions. For a given inflow speed, the
higher is the magnetic energy stored in the region, the smaller is the
reconnection rate.

If we assume that the reconnection rate is between 10$^{-3}$ and 10$^{-1}$
\citep[see e.g.,][]{nar06}, we can infer the inflow speed required to trigger
the reconnection. The results for the four studied active regions are summarized
in Fig.~\ref{fig:inflow}, where we plot the inflow speed as a function of height
for different reconnection rates ($M_e= 0.1, 0.01, 0.001$) for both potential
(dashed curves) and \nlff~(solid curves) fields. The variation with height of
the inflow  speed is consistent with the variation of the Alfv\'en speed as
described in Fig.~\ref{fig:va_iso}. From these curves, we can deduce a threshold
on the reconnection rate in order to obtain inflow speeds in agreement with
observations \citep[e.g.,][]{nar06, nag06}. For instance, if a reconnection
process occurs in AR 8151, the reconnection rate should be of the order or below
0.01 which gives typical inflow speeds below 50 \kms.

We also note that the minimum of the inflow speed or Alfv\'en speed (see
Section~\ref{sec:model_param}) is the location in the low corona where the
reconnection process is more likely to happen. The conclusion is valid in a
statistical sense because we only consider average values of the inflow speed
and because, in order for it to occur, the reconnection process needs a
particular magnetic topology and a diffusion region which are beyond the scope
of this article.

	\subsection{Coronal plasma $\beta$}

An important quantity in plasma physics is the plasma $\beta$, the ratio of the
gas pressure to the magnetic pressure. When $\beta \ll 1$, the plasma is
dominated by the magnetic field. Nevertheless, the plasma $\beta$ in the solar
atmosphere varies greatly with height \citep{gar01}. In addition, we note that
the plasma $\beta$ varies from one structure to another and even coronal
structures, such as filaments, can have $\beta \geq 1$. As mentioned in
Section~\ref{sec:model_param}, the plasma $\beta$ can be expressed as follows:
\begin{equation} 
\beta = \frac{2~c_s^2}{\gamma v_A^2} 
\end{equation} 
for an isothermal atmosphere ($\gamma = 1$). In Fig.~\ref{fig:beta}, we plot the
variation of the average plasma $\beta$ with height for the four active regions
described in Section~\ref{sec:obs_ars}. For both potential and \nlff~models, the
plasma $\beta$ is less than 1. This result is consistent with the typical nature
of the coronal active region magnetic field. The plasma $\beta$ is larger for
the potential field because the average Alfv\'en speed decreases with height
more rapidly than for the \nlff~field. The maximum of the plasma $\beta$ is
located at the same height as the minimum of Alfv\'en speed (see
Section~\ref{sec:model_va}) and also  satisfies Eqn.~(\ref{eq:min_va}). The
plasma $\beta$ tends to zero for a constant gravitational field, whilst it
increases towards 1 for the varying gravity model. The latter is then in
agreement with plasma $\beta$ measurements reported by \cite{gar01}.

	\subsection{Average Alfv\'en transit time}
	
The Alfv\'en transit time is an important quantity for the propagation of Alfv\'en
waves and for the relaxation of a magnetic configuration. In this study of
global properties of Alfv\'en speeds, we derive the average Alfv\'en transit
time associated with two different lengths:
\begin{equation}
\tau_{A}^{(H)} = \frac{H}{v_A}
\end{equation}
and
\begin{equation}
\tau_{A}^{(\pi L)} = \frac{\pi L}{v_A},
\end{equation}
where $v_A$ is the Alfv\'en speed, $H$ is the pressure scale-height and $L$ is
the length of the computational box. We suppose $\tau_{A}^{(H)}$ (resp. 
$\tau_{A}^{(\pi L)}$) is a minimum (resp. maximum) of the Alfv\'en transit time.
The larger the Alfv\'en transit time is, the more stable the equilibrium
is. In Fig.~\ref{fig:tau}, we plot the different Alfv\'en transit times for the
four active regions. At the base of the corona (5 Mm), the maximum Alfv\'en time
is 750s for AR 8151, 3600s for AR 8210, 70s for AR 9077, 300s for AR 10486.
These results mean that the most stable active  region is AR 8210 associated
with confined C-class flares, and the least stable one is AR 9077 exhibiting
post-flare loops. It is worth noticing that the average Alfv\'en transit time is
an interesting quantity only when discussing the global or statistical
properties of  magnetic configurations. A more appropriate quantity is the
Alfv\'en transit time along a single field line which is beyond the scope of
this article. Of importance for MHD modelling, the Alfv\'en transit time along
field lines is not a constant with height which makes it difficult to estimate
the scaling of the time evolution with this quantity.

Despite the simple assumptions of this model, we have derived interesting
properties of the Alfv\'en speed in the solar corona. In a forthcoming paper, we
will focus on the properties of individual field lines.

\begin{acknowledgements}
We would like to thank the referee for his useful comments which helped to
improve the article. We also would like to thank L. Fletcher, H. Hudson and S.
Galtier for interesting discussions. We thank the UK STFC for financial support
(STFC RG). The computations have been done using the XTRAPOL code developed by
T. Amari (Ecole Polytechnique, France). We also acknowledge the financial
support by the European Commission through the SOLAIRE network
(MTRN-CT-2006-035484).
\end{acknowledgements}

\bibliographystyle{aa}
\bibliography{/local_raid/stephane/TEX/Bib/mybib}

\end{document}